\documentclass[a4paper,fleqn,usenatbib,onecolumn]{mnras}
\usepackage[T1]{fontenc}
\usepackage{aecompl}
\usepackage{graphicx}
\usepackage{epstopdf}
\usepackage{amsmath}
\usepackage{amssymb}	
\usepackage{enumerate}

\title[Analyzing BAO in Sparse Spectroscopic Samples]{Analyzing Baryon Acoustic Oscillations in Sparse Spectroscopic Samples via Cross-Correlation with Dense Photometry}
\author[A. Patej \& D. J. Eisenstein]{
Anna Patej$^{1,2}$ and 
Daniel J. Eisenstein$^{3}$ \\
$^{1}$Einstein Fellow\\
$^{2}$Steward Observatory, University of Arizona, 933 N. Cherry Ave., Tucson, AZ 85719, USA\\
$^{3}$Harvard-Smithsonian Center for Astrophysics, 60 Garden St., Cambridge, MA 02138, USA
}

\begin{document} 

\pagerange{\pageref{firstpage}--\pageref{lastpage}}\pubyear{2015}
\maketitle
\label{firstpage}

\begin{abstract}
We develop a formalism for measuring the cosmological distance scale from baryon acoustic oscillations (BAO) using the cross-correlation of a sparse redshift survey with a denser photometric sample. This reduces the shot noise that would otherwise affect the auto-correlation of the sparse spectroscopic map. As a proof of principle, we make the first on-sky application of this method to a sparse sample defined as the $z>0.6$ tail of the Sloan Digital Sky Survey's (SDSS) BOSS/CMASS sample of galaxies and a dense photometric sample from SDSS DR9. We find a $2.8\sigma$ preference for the BAO peak in the cross-correlation at an effective $z=0.64$, from which we measure the angular diameter distance $D_{M}(z=0.64) = (2418\pm 73\;\mathrm{Mpc}) \left(r_s/r_{s,\mathrm{fid}}\right)$. Accordingly, we expect that using this method to combine sparse spectroscopy with the deep, high quality imaging that is just now becoming available will enable higher precision BAO measurements than possible with the spectroscopy alone. 
\end{abstract}

\section{Introduction}\label{s:intro}
The signature of baryon acoustic oscillations (BAO) is a peak in the clustering of galaxies at large scales that is a relic of propagating sound waves in the early universe and whose location depends on the distance the waves were able to cover prior to recombination~\citep[for a recent review, see][]{weinberg13}. Major redshift surveys like the Sloan Digital Sky Survey's (SDSS) Baryon Oscillation Spectroscopic Survey~\citep[BOSS;][]{dawson13} and the 2dF Galaxy Redshift Survey~\citep[2dFGRS;][]{colless01} have analyzed the clustering of galaxies in order to detect this feature and use it to measure the cosmological quantities $D_M(z)$, the angular diameter distance, and $H(z)$, the Hubble parameter \citep[][]{eisenstein05,cole05}. BOSS has already attained percent level precision distance measurements using the BAO method at $z=0.57$~\citep{anderson14}.

However, most detections of the BAO feature in samples of galaxies thus far have occurred at fairly low redshifts, where these galaxy redshift surveys are able to attain sufficiently high densities of galaxies with spectroscopic redshift measurements. New measurements of quasar clustering have recently enabled BAO detections to higher redshifts \citep{ata17} using eBOSS \citep{dawson16}, while future surveys such as the Dark Energy Spectroscopic Instrument (DESI) \citep{levi13} will push the availability of spectra at higher redshifts by targeting quasars in addition to lower redshift galaxies, for which new approaches will be required to fully leverage these sparser data.

To this end, in this paper we develop the formalism of a method for measuring the BAO signal with a sparse sample of galaxies with spectroscopic redshifts that consists of cross-correlating the sparse data with a dense photometric sample to enhance the signal. Our work is related to concepts previously introduced by \citet{nishizawa13}, but we more fully develop the configuration-space statistic, deriving explicit forms for the correlation function and the covariance matrix. We then make the first application of this method to data from SDSS to verify whether we can detect the BAO peak. In what follows, we use a flat $\Lambda$CDM cosmology with $\Omega_{m} = 0.314$ and $h=0.67$, similar to \citet{alam17}.

\section{Motivation}
We begin with a pedagogical motivation of when employing the cross-correlation is useful, with a more detailed calculation to follow in Section~\ref{s:mf}. As we are interested in large physical scales, for this section and the remainder of the paper, we assume that the galaxy density field is described by a Gaussian random field and a linear bias model. For a spectroscopic survey of galaxy density $n_s$, the variance on each Fourier mode is $\sigma_s^2 = (P_s + 1/n_s)^2$, so that the signal to noise $\mathrm{(S/N)}$ is given by \citep[][]{weinberg13}:
\begin{align}\label{e:sn}
\frac{S}{N} = \frac{n_sP_s}{1+n_sP_s},
\end{align}
where $P_s$ denotes the power spectrum $P_s(k)$. From this, we immediately see that the S/N per mode approaches 1 when $n_sP_s \gg 1$ but decreases when $n_sP_s < 1$, so that the relevant quantity to track is $nP$.

Let us now assume we have a photometric dataset with some density $n_p$ and power spectrum $P_p$. Both the spectroscopic and photometric samples are biased tracers of the underlying matter distribution such that:
\begin{align}
\delta_s(\boldsymbol{x}) = b_s \delta(\boldsymbol{x}), \; \delta_p(\boldsymbol{x}) = b_p \delta(\boldsymbol{x})
\end{align}
Accordingly, we may write the power spectra as $P_s (k)= b_s^2P(k)$ and $P_p(k) = b_p^2(n_p/N_p)P(k)$, with the factor of $n_p/N_p = 1/L_p$, where $L_p$ is the line-of-sight spatial depth of the photometry, approximately accounting for the fact that we are converting the three-dimensional $P(k)$ to the two-dimensional $P_p(k)$.

When we cross-correlate a photometric sample to the spectroscopic sample, we obtain the power spectrum of the cross-correlation as $P_x = (n_p/N_p)b_sb_pP$ with variance $\sigma_x^2 =  P_x^2+\sigma_{s,2}\sigma_{p,2}$, where the subscript ``$2$'' indicates projection, so that the signal-to-noise becomes:
\begin{align}
\left(\frac{S}{N}\right)^2 &= \frac{P_x^2}{P_x^2+(b_s^2[n_s/N_s]P_s+1/N_s)(b_p^2[n_p/N_p]P_p+1/N_p)},\\
\left(\frac{S}{N}\right)^2 &= \left(1+\frac{n_s}{N_s}\frac{N_p}{n_p}\left[1+\frac{1}{n_sb_s^2P}\right]\left[1+\frac{1}{n_pb_p^2P}\right]\right)^{-1},\\
\left(\frac{S}{N}\right)^2 &= \left(1+\frac{L_p}{L_s}\left[1+\frac{1}{n_sb_s^2P}\right]\left[1+\frac{1}{n_pb_p^2P}\right]\right)^{-1}.\label{e:sn_cross}
\end{align}
Consequently, we see that with the cross-correlation, when $n_sP_s$ gets small, we can, to a certain extent, compensate with the $n_pP_p$ term. More specifically, while Equation~\ref{e:sn} goes as $n_s$ for small $n_s$, Equation~\ref{e:sn_cross} goes as $n_s^{1/2}$.

To see when this method would improve upon the results of using the auto-correlation, let us estimate the number of modes that we could measure. For a spectroscopic sample with constant density, the effective number of modes $\mathcal{N}_s$ in a shell of width $\Delta k$ is given by:
\begin{align}
\mathcal{N}_s = \frac{AL_s}{2(2\pi)^3}4\pi k^2 \Delta k\times \left(\frac{n_sP_s}{1+n_sP_s}\right)^2.
\end{align}
In the case of the photometric sample, we have $k_z = 0$, which means that 
\begin{align}
\mathcal{N}_p = \frac{A}{(2\pi)^2}2\pi k \Delta k\times \left(1+\frac{L_p}{L_s}\left[1+\frac{1}{n_sb_s^2P}\right]\left[1+\frac{1}{n_pb_p^2P}\right]\right)^{-1}.
\end{align}
We are looking for the case wherein $\mathcal{R} = \mathcal{N}_s/\mathcal{N}_p < 1$, which means that
\begin{align}
\mathcal{R} = \frac{kL_s}{2\pi}\left(\frac{1}{1+1/(n_sb_s^2P)}\right)^2\times \left(1+\frac{L_p}{L_s}\left[1+\frac{1}{n_sb_s^2P}\right]\left[1+\frac{1}{n_pb_p^2P}\right]\right),
\end{align}
and if we assume that $n_sb_s^2P \ll 1$ (the limit in which the auto-correlation method begins to lose information), this simplifies to
\begin{align}
\mathcal{R} &= \frac{kL_s}{2\pi}\left(n_sb_s^2P\right)^{2}\times \left(\frac{L_p}{L_s}\left[n_sb_s^2P\right]^{-1}\left[1+\frac{1}{n_pb_p^2P}\right]\right),\\
\mathcal{R} &=\frac{kL_p}{2\pi}\left(\frac{1+1/(n_pb_p^2P)}{1/(n_sb_s^2P)}\right).
\end{align}
Let us denote $\tau_s = n_sb_s^2P$ and $\tau_p = n_pb_p^2P$, so that:
\begin{align}
\mathcal{R} &= \frac{kL_p}{2\pi}\frac{1+\tau_p^{-1}}{\tau_s^{-1}}.
\end{align}
To satisfy $\mathcal{R}<1$, we require 
\begin{align}
\tau_s^{-1} > \frac{kL_p}{2\pi}\left(1+\tau_p^{-1}\right).
\end{align}
We can usually get photometric samples of high density, corresponding to the limit $\tau_p^{-1}\rightarrow0$. Accordingly, for a reasonable photometric depth of $L_p = 500 \;h^{-1}\mathrm{Mpc}$ and $k = 0.15 \; h\mathrm{Mpc}^{-1}$, we require $\tau_s \lesssim 0.08$. We additionally note that this method can be beneficial even when the condition that $\tau_s \lesssim 0.08$ is \textit{not} satisfied; shot noise always degrades the BAO precision, particularly when $\tau_s<1$. In that case, even though the auto-correlation will give a more precise distance, the cross-correlation is still adding some new information since the individual modes are not being well measured by the auto-correlation.

\section{Mathematical Formalism}\label{s:mf}
\subsection{Framework}
We now more formally derive our method. Let us assume that we have two samples of galaxies: a spectroscopic sample with fractional overdensity field $\delta_s(\boldsymbol{r})$, and the other a photometric sample. For this latter sample, we can calculate a projected overdensity field as
\begin{align}
\Delta_p(\boldsymbol{\theta}) &= \frac{\int dr^{\prime}  \:r^{\prime2} n_p(r^{\prime})\delta_p(\boldsymbol{r}^{\prime})}{\int dr^{\prime} \:  r^{\prime2}n_p(r^{\prime})}.
\end{align}
Defining the normalization factor as
\begin{align}
\mathcal{N}\equiv \int dr^{\prime}\: r^{\prime2} n_p(r^{\prime}),
\end{align}
this may be written simply as
\begin{align}\label{e:Deltap}
\Delta_p(\boldsymbol{\theta}) &= \frac{1}{\mathcal{N}}\int dr^{\prime}  \:r^{\prime2} n_p(r^{\prime})\delta_p(\boldsymbol{r}^{\prime}).
\end{align}

The correlation of $\delta_s$ and $\Delta_p$, noting that $\boldsymbol{r}=(r,\boldsymbol{\theta})$, is then
\begin{align}
W(\boldsymbol{R}) &= \langle \delta_s(\boldsymbol{r})\Delta_p(\boldsymbol{\theta}+\boldsymbol{R}/r)\rangle,\\
W(\boldsymbol{R}) &= \frac{\int d^3r  \: \delta_s(\boldsymbol{r})½n_s(r)\mathcal{W}(r,\eta)\int d^2\theta^{\prime} \:\Delta_p(\boldsymbol{\theta^{\prime}})\mathcal{N}\mathcal{S}(\boldsymbol{\theta},\boldsymbol{\theta}^{\prime},\boldsymbol{R},r)}{\int d^3r  \: n_s(r)\mathcal{W}(r,\eta)\int d^2\theta^{\prime} \:\mathcal{N}\mathcal{S}(\boldsymbol{\theta},\boldsymbol{\theta}^{\prime},\boldsymbol{R},r)}.\label{e:w_def}
\end{align}
Here, $\mathcal{W}$ is a weighting function that may depend on $r$ and other variables independent of $\boldsymbol{r}$, collectively referred to as $\eta$, and $\mathcal{S}$ is a separation function that depends on the angular coordinates of the galaxies. If we focus on a single separation, then the separation function is a Dirac delta function:
\begin{align}\label{e:selfunc}
\mathcal{S}(\boldsymbol{\theta},\boldsymbol{\theta}^{\prime},\boldsymbol{R},r) = \delta_D^{(2)}\left(r\boldsymbol{\theta}^{\prime}-\left(r\boldsymbol{\theta}+\boldsymbol{R}\right)\right).
\end{align}
Using Equations \ref{e:Deltap} and \ref{e:selfunc} and writing out explicitly $\boldsymbol{r} = (r,\boldsymbol{\theta})$, Equation \ref{e:w_def} becomes
\begin{align}
W(\boldsymbol{R}) &= \frac{\int \! dr \:r^2n_s(r)\mathcal{W}(r,\eta) \!\int\! d^2\theta  \: \delta_s(r,\boldsymbol{\theta})\!\int \! d^2\theta^{\prime} \: \delta_D^{(2)}\!\left(r\boldsymbol{\theta}^{\prime}\! -\!\left(r\boldsymbol{\theta}\! +\!\boldsymbol{R}\right)\right)\!\int \! dr^{\prime}  \:r^{\prime2}n_p(r^{\prime})\delta_p(r^{\prime},\boldsymbol{\theta}^{\prime})}{\int dr \:r^2n_s(r)\mathcal{W}(r,\eta)\int d^2\theta \:\int d^2\theta^{\prime} \: \delta_D^{(2)}\!\left(r\boldsymbol{\theta}^{\prime}\! -\!\left(r\boldsymbol{\theta}\! +\!\boldsymbol{R}\right)\right)\int dr^{\prime}  \:r^{\prime2}n_p(r^{\prime})}.
\end{align}
The delta function collapses the $\theta^{\prime}$ integral in both the numerator and denominator, giving $1/r^2$, and in the denominator the $\theta$ integral can be immediately evaluated to a constant, yielding
\begin{align}
W(\boldsymbol{R}) &= \frac{\int dr \:n_s(r)\mathcal{W}(r,\eta) \int d^2\theta  \: \delta_s(r,\boldsymbol{\theta})\int dr^{\prime}r^{\prime2}  \:n_p(r^{\prime})\delta_p(r^{\prime},\boldsymbol{\theta}+\boldsymbol{R}/r)}{4\pi\int dr  \: n_s(r)\mathcal{W}(r,\eta)\int dr^{\prime}  \:r^{\prime2}n_p(r^{\prime})}.
\end{align}
Rearranging the order of integration gives
\begin{align}\label{e:w_final_int}
W(\boldsymbol{R}) &= \frac{\int dr \:n_s(r)\mathcal{W}(r,\eta) \int dr^{\prime}  \:r^{\prime2}n_p(r^{\prime}) \int d^2\theta  \: \delta_s(r,\boldsymbol{\theta})\delta_p(r^{\prime},\boldsymbol{\theta}+\boldsymbol{R}/r)}{4\pi\int dr  \: n_s(r)\mathcal{W}(r,\eta)\int dr^{\prime}  \:r^{\prime2}n_p(r^{\prime})}.
\end{align}

With an eye towards an application to catalog data, we note that the correlation function is constructed from pair counts of galaxies, normalized by pair counts of random galaxies, so that
\begin{align}
W(\boldsymbol{R}) = \frac{N_sN_p}{R_sR_p}.
\end{align}
This allows us to identify the numerator and denominator of Equation~\ref{e:w_final_int} with $N_sN_p$ and $R_sR_p$, where both terms are a function of $R$, the separation between the galaxies in the pair. 

\subsection{Expectation Value}
Taking the expectation value of Equation \ref{e:w_final_int} yields
\begin{align}
\langle W(\boldsymbol{R}) \rangle &= \frac{\int dr \:n_s(r)\mathcal{W}(r,\eta) \int dr^{\prime} \:r^{\prime2}n_p(r^{\prime}) \int d^2\theta  \: \langle\delta_s(r,\boldsymbol{\theta})\delta_p(r^{\prime},\boldsymbol{\theta}+\boldsymbol{R}/r)\rangle}{4\pi\int dr\: n_s(r)\mathcal{W}(r,\eta)\int dr^{\prime} \: r^{\prime2}n_p(r^{\prime})}.
\end{align}
Focusing on the numerator, 
\begin{align}
\langle N_sN_p(\boldsymbol{R}) \rangle = \int dr \: n_s(r)\mathcal{W}(r,\eta) \int dr^{\prime}  \:r^{\prime2}n_p(r^{\prime}) \int d^2\theta  \: \langle\delta_s(r,\boldsymbol{\theta})\delta_p(r^{\prime},\boldsymbol{\theta}+\boldsymbol{R}/r)\rangle,
\end{align}
we can use the Fourier transform,
\begin{align}
\delta(\boldsymbol{r})=\frac{1}{(2\pi)^{3/2}}\int d^3k\:\delta_{\boldsymbol{k}}e^{i\boldsymbol{k}\cdot \boldsymbol{r}}, 
\end{align}
as well as the formalism of \citet{dekel99}, which relates the $\delta_g = \delta_s,\delta_p$ fields to the underlying matter density field $\delta\equiv\delta_m$ as
\begin{align}
\delta_g = b_g\delta+\epsilon_g,
\end{align}
to write the expectation value of two $\delta$s as
\begin{align}
\langle\delta(r,\boldsymbol{\theta})\delta(r^{\prime},\boldsymbol{\theta}+\boldsymbol{R}/r)\rangle = \frac{b_sb_p}{(2\pi)^3}\int\int d^3kd^3k^{\prime}\: \langle\delta_{\boldsymbol{k}}\delta_{\boldsymbol{k}^{\prime}}\rangle e^{ir\boldsymbol{k}_{\perp}\cdot\boldsymbol{\theta}}e^{irk_{\shortparallel}}e^{ir^{\prime}\boldsymbol{k}^{\prime}_{\perp}\cdot(\boldsymbol{\theta}+\boldsymbol{R}/r)}e^{ir^{\prime}k_{\shortparallel}^{\prime}},
\end{align}
where we are using the flat sky approximation and have written $k = \sqrt{k_{\perp}^2+k_{\shortparallel}^2}$, with $k_{\perp}$ indicating the transverse components of $k$ and $k_{\shortparallel}$ indicating the line-of-sight component. Then, using
\begin{align}
\langle\delta_{\boldsymbol{k}}\delta_{\boldsymbol{k}^{\prime}}\rangle = \delta_D^{(3)}(\boldsymbol{k}+\boldsymbol{k}^{\prime})P(k),
\end{align}
and integrating over $k$ yields
\begin{align}
\langle\delta(r,\boldsymbol{\theta})\delta(r^{\prime},\boldsymbol{\theta}+\boldsymbol{R}/r)\rangle &= \frac{b_sb_p}{(2\pi)^3}\int d^3k^{\prime}\: P(k^{\prime}) e^{-ir\boldsymbol{k}_{\perp}^{\prime}\cdot\boldsymbol{\theta}}e^{-irk_{\shortparallel}}e^{ir^{\prime}\boldsymbol{k}_{\perp}^{\prime}\cdot(\boldsymbol{\theta}+\boldsymbol{R}/r)}e^{ir^{\prime}k_{\shortparallel}^{\prime}},\\
 &= \frac{b_sb_p}{(2\pi)^3}\int d^3k^{\prime}\: P(k^{\prime}) e^{-ir\boldsymbol{k}_{\perp}^{\prime}\cdot\boldsymbol{\theta}}e^{ir^{\prime}\boldsymbol{k}_{\perp}^{\prime}\cdot(\boldsymbol{\theta}+\boldsymbol{R}/r)}e^{-i(r-r^{\prime})k_{\shortparallel}^{\prime}}.
\end{align}
Thus, 
\begin{align}
\langle N_sN_p \rangle =\;\; &b_sb_p\int dr \:n_s(r)\mathcal{W}(r,\eta) \int dr^{\prime}  \:r^{\prime2}n_p(r^{\prime}) \int \frac{d^2\theta}{(2\pi)^3} \int d^3k^{\prime}\: P(k^{\prime}) e^{-ir\boldsymbol{k}_{\perp}^{\prime}\cdot\boldsymbol{\theta}}e^{ir^{\prime}\boldsymbol{k}_{\perp}^{\prime}\cdot(\boldsymbol{\theta}+\boldsymbol{R}/r)}e^{-i(r-r^{\prime})k_{\shortparallel}^{\prime}},
\end{align}
If we assume that the line-of-sight weighting is much larger than the BAO scale, then we may approximate $k = \sqrt{k_{\perp}^2+k_{\shortparallel}^2} \approx k_{\perp}$ and split up the last integral:
\begin{align}
\langle N_sN_p \rangle &= b_sb_p\int dr \:n_s(r)\mathcal{W}(r,\eta) \int dr^{\prime}  \:r^{\prime2}n_p(r^{\prime}) \int \frac{d^2\theta}{(2\pi)^3}\int d^2k_{\perp}^{\prime}\: P(k_{\perp}^{\prime}) e^{-ir\boldsymbol{k}_{\perp}^{\prime}\cdot\boldsymbol{\theta}}e^{ir^{\prime}\boldsymbol{k}_{\perp}^{\prime}\cdot(\boldsymbol{\theta}+\boldsymbol{R}/r)}\int dk_{\shortparallel}^{\prime} \; e^{-i(r-r^{\prime})k_{\shortparallel}^{\prime}},\\
&= b_sb_p\int dr \:n_s(r)\mathcal{W}(r,\eta) \int dr^{\prime}  \:r^{\prime2}n_p(r^{\prime}) \int \frac{d^2\theta}{(2\pi)^3}\int d^2k_{\perp}^{\prime}\: P(k_{\perp}^{\prime}) e^{-ir\boldsymbol{k}_{\perp}^{\prime}\cdot\boldsymbol{\theta}}e^{ir^{\prime}\boldsymbol{k}_{\perp}^{\prime}\cdot(\boldsymbol{\theta}+\boldsymbol{R}/r)}2\pi\delta_D(r-r^{\prime}),\\
&=b_sb_p\int dr \:r^2n_s(r)\mathcal{W}(r,\eta)n_p(r) \int \frac{d^2\theta}{(2\pi)^2}  \int d^2k_{\perp}^{\prime}\: P(k_{\perp}^{\prime}) e^{-ir\boldsymbol{k}_{\perp}^{\prime}\cdot\boldsymbol{\theta}}e^{ir\boldsymbol{k}_{\perp}^{\prime}\cdot(\boldsymbol{\theta}+\boldsymbol{R}/r)}.
\end{align}
Now, the $\boldsymbol{\theta}$ terms cancel, making the integral over that variable trivial:
\begin{align}
\langle N_sN_p \rangle &=b_sb_p \int dr \:r^2n_s(r)\mathcal{W}(r,\eta)n_p(r) \;\frac{4\pi}{(2\pi)^2}  \int d^2k_{\perp}^{\prime}\: P(k_{\perp}^{\prime}) e^{ir\boldsymbol{k}_{\perp}^{\prime}\cdot\boldsymbol{R}/r},\\
&= \frac{b_sb_p}{\pi} \int dr \:r^2n_s(r)\mathcal{W}(r,\eta)n_p(r)  \int d^2k_{\perp}^{\prime}\: P(k_{\perp}^{\prime}) e^{i\boldsymbol{k}_{\perp}^{\prime}\cdot\boldsymbol{R}}.
\end{align}
We can then identify:
\begin{align}
w_p(\boldsymbol{R}) \equiv \frac{1}{2\pi} \int d^2k_{\perp}^{\prime}\: P(k_{\perp}^{\prime}) e^{i\boldsymbol{k}_{\perp}^{\prime}\cdot\boldsymbol{R}},
\end{align}
where $w_p$ is the projected correlation function \citep{davis83}, which has been used extensively in galaxy clustering analyses \citep[e.g.,][]{coil08,zehavi11,nuza13} as it is less affected by redshift space distortions. The function $w_p$ can be calculated as the line-of-sight integral of the three-dimensional correlation function,
\begin{align}
w_p(r_{\perp}) = \int_{-\infty}^{\infty} dr_{\shortparallel}\:\xi(r_{\perp},r_{\shortparallel}).
\end{align}
Returning to our derivation, we find
\begin{align}
\langle N_sN_p \rangle &=2 b_sb_p\int dr \:r^2n_s(r)\mathcal{W}(r,\eta)n_p(r) w_p(\boldsymbol{R}).
\end{align}
Finally, this yields
\begin{align}\label{e:final_w}
\langle W(\boldsymbol{R}) \rangle &= \frac{b_sb_p}{2\pi}w_p(\boldsymbol{R})\frac{\int dr \:r^2n_s(r)\mathcal{W}(r,\eta)n_p(r) }{\int dr  \: n_s(r)\mathcal{W}(r,\eta)\int dr^{\prime}  \:r^{\prime2}n_p(r^{\prime})}.
\end{align}

To gain some intuition for the behavior of $\langle w \rangle$, we assume that the distributions of the spectroscopic and photometric galaxies are given by boxcar functions of lengths $L_s$ and $L_p$, respectively, which represent the depths of the survey, and that $L_s < L_p$. Let us also assume the weighting is independent of $r$, and equal to 1. In that case, the integral normalization gives
\begin{align}
\frac{\int dr \:r^2n_s(r)\mathcal{W}(r,\eta)n_p(r) }{\int dr  \: n_s(r)\mathcal{W}(r,\eta)\int dr^{\prime}  \:r^{\prime2}n_p(r^{\prime})} = \frac{\int_{0}^{L_s}dr \:r^2}{\int_{0}^{L_s}dr \int_{0}^{L_p}dr' \:r'^2} = \frac{L_s^3}{L_p^3L_s} = \frac{L_s^2}{L_p^3}\sim\frac{1}{L}.
\end{align}
Accordingly, we see that this factor roughly scales as the inverse of the depth of the survey.

\subsection{Covariance}
The covariance of $W$ is given by
\begin{align}
c(\boldsymbol{R},\boldsymbol{S}) &\equiv \langle[W(\boldsymbol{R})-\langle W(\boldsymbol{R})\rangle][W(\boldsymbol{S})-\langle W(\boldsymbol{S})\rangle]\rangle = \langle W(\boldsymbol{R})W(\boldsymbol{S})\rangle -\langle W(\boldsymbol{R})\rangle\langle W(\boldsymbol{S})\rangle.\label{e:cov}
\end{align}
Recalling the fractional form of $W$ from Equation~\ref{e:final_w}, the numerator of first term in Equation (\ref{e:cov}), which we will refer to as $N_C$ for simplicity (and denote the denominator by $D_C$), may be written out as
\begin{align}
\langle N_C\rangle =& D_C\langle W(\boldsymbol{R})W(\boldsymbol{S}) \rangle \\
=&\int dr_1\: n_s(r_1)\mathcal{W}(r_1,\eta)\int dr_1^{\prime}\:r_1^{\prime2}n_p(r_1^{\prime})\int dr_2\: n_s(r_2)\mathcal{W}(r_2,\eta)\int dr_2^{\prime}\:r_2^{\prime2} n_p(r_2^{\prime})\int d^2\theta_1\nonumber\\
&\times \int d^2\theta_2\langle\delta_s(r_1,\boldsymbol{\theta}_1)\delta_p(r_1^{\prime},\boldsymbol{\theta}_1+\boldsymbol{R}/r_1)\delta_s(r_2,\boldsymbol{\theta}_2)\delta_p(r_2^{\prime},\boldsymbol{\theta}_2+\boldsymbol{S}/r_2)\rangle.
\end{align}
Again using the \citet{dekel99} formalism so that $\delta_g = b_g\delta+\epsilon_g$, this immediately becomes
\begin{align}\label{e:cov_num}
\langle N_C\rangle =  \; &b_s^2b_p^2\int dr_1\:n_s(r_1)\mathcal{W}(r_1,\eta)\int dr_1^{\prime}\:r_1^{\prime2}n_p(r_1^{\prime})\int dr_2\:  n_s(r_2)\mathcal{W}(r_2,\eta)\int dr_2^{\prime}\: r_2^{\prime2}n_p(r_2^{\prime})\int d^2\theta_1\nonumber\\
&\times \int d^2\theta_2\langle\delta(r_1,\boldsymbol{\theta}_1)\delta(r_1^{\prime},\boldsymbol{\theta}_1+\boldsymbol{R}/r_1)\delta(r_2,\boldsymbol{\theta}_2)\delta(r_2^{\prime},\boldsymbol{\theta}_2+\boldsymbol{S}/r_2)\rangle.
\end{align}
The expectation value may be expressed as a Fourier transform:
\begin{align}
\langle\delta(r_1,\boldsymbol{\theta}_1)\delta(r_1^{\prime},\boldsymbol{\theta}_1+\boldsymbol{R}/r_1)\delta(r_2,\boldsymbol{\theta}_2)\delta(r_2^{\prime},\boldsymbol{\theta}_2+\boldsymbol{S}/r_2)\rangle = & \frac{1}{(2\pi)^6}\int\int\int\int d^3kd^3k^{\prime}d^3qd^3q^{\prime}\langle\delta_{\boldsymbol{k}}\delta_{\boldsymbol{k}^{\prime}}\delta_{\boldsymbol{q}}\delta_{\boldsymbol{q}^{\prime}}\rangle e^{ir_1\boldsymbol{k}_{\perp}\cdot\boldsymbol{\theta}_1}e^{ir_1^{\prime}\boldsymbol{k}_{\perp}^{\prime}\cdot(\boldsymbol{\theta}_1+\boldsymbol{R}/r_1)}\nonumber\\
&\times e^{ir_2\boldsymbol{q}_{\perp}\cdot\boldsymbol{\theta_2}}e^{ir_2^{\prime}\boldsymbol{q}_{\perp}^{\prime}\cdot(\boldsymbol{\theta}_2+\boldsymbol{S}/r_2)}e^{ir_1k_{\shortparallel}}e^{ir_1^{\prime}k_{\shortparallel}^{\prime}}e^{ir_2q_{\shortparallel}}e^{ir_2^{\prime}q_{\shortparallel}^{\prime}}.
\end{align}
Assuming a Gaussian random field and using Wick's theorem, we can write the expectation value as
\begin{align}
\langle\delta_{\boldsymbol{k}}\delta_{\boldsymbol{k}^{\prime}}\delta_{\boldsymbol{q}}\delta_{\boldsymbol{q}^{\prime}}\rangle = \langle \delta_{\boldsymbol{k}}\delta_{\boldsymbol{k}^{\prime}}\rangle\langle\delta_{\boldsymbol{q}}\delta_{\boldsymbol{q}^{\prime}}\rangle +\langle\delta_{\boldsymbol{k}}\delta_{\boldsymbol{q}}\rangle\langle\delta_{\boldsymbol{k}^{\prime}}\delta_{\boldsymbol{q}^{\prime}}\rangle+\langle \delta_{\boldsymbol{k}}\delta_{\boldsymbol{q}^{\prime}}\rangle\langle \delta_{\boldsymbol{k}^{\prime}}\delta_{\boldsymbol{q}}\rangle.
\end{align}
The first term in this expansion can immediately be identified with $\langle w(\boldsymbol{R})\rangle\langle w(\boldsymbol{S})\rangle$, which cancels the second term in Equation (\ref{e:cov}). Then, we express the remaining factors in terms of the power spectrum:
\begin{align}
\langle\delta_{\boldsymbol{k}}\delta_{\boldsymbol{q}}\rangle\langle\delta_{\boldsymbol{k}^{\prime}}\delta_{\boldsymbol{q}^{\prime}}\rangle &= \delta^{(3)}_D(\boldsymbol{k}+\boldsymbol{q})P(q) \delta^{(3)}_D(\boldsymbol{k}^{\prime}+\boldsymbol{q}^{\prime})P(q^{\prime}),\\
\langle \delta_{\boldsymbol{k}}\delta_{\boldsymbol{q}^{\prime}}\rangle\langle \delta_{\boldsymbol{k}^{\prime}}\delta_{\boldsymbol{q}}\rangle &= \delta^{(3)}_D(\boldsymbol{k}+\boldsymbol{q}^{\prime})P(q^{\prime}) \delta^{(3)}_D(\boldsymbol{k}^{\prime}+\boldsymbol{q})P(q).
\end{align}
Let us denote the Fourier transform of the first of these terms by $A$ and that of the second by $B$.

Beginning with term $A$, we find
\begin{align}
A =  &\;\frac{1}{(2\pi)^6}\int\int\int\int d^3kd^3k^{\prime}d^3qd^3q^{\prime}\:\langle\delta_{\boldsymbol{k}}\delta_{\boldsymbol{q}}\rangle\langle\delta_{\boldsymbol{k}^{\prime}}\delta_{\boldsymbol{q}^{\prime}}\rangle e^{ir_1\boldsymbol{k}_{\perp}\cdot\boldsymbol{\theta}_1}e^{ir_1^{\prime}\boldsymbol{k}_{\perp}^{\prime}\cdot(\boldsymbol{\theta}_1+\boldsymbol{R}/r_1)}e^{ir_2\boldsymbol{q}_{\perp}\cdot\boldsymbol{\theta_2}}e^{ir_2^{\prime}\boldsymbol{q}_{\perp}^{\prime}\cdot(\boldsymbol{\theta}_2+\boldsymbol{S}/r_2)}e^{ir_1k_{\shortparallel}}e^{ir_1^{\prime}k_{\shortparallel}^{\prime}}e^{ir_2q_{\shortparallel}}e^{ir_2^{\prime}q_{\shortparallel}^{\prime}},\\
=  &\;\frac{1}{(2\pi)^6}\int\int\int\int d^3kd^3k^{\prime}d^3qd^3q^{\prime}\:\delta^{(3)}_D(\boldsymbol{k}+\boldsymbol{q})P(q) \delta^{(3)}_D(\boldsymbol{k}^{\prime}+\boldsymbol{q}^{\prime})P(q^{\prime})e^{ir_1\boldsymbol{k}_{\perp}\cdot\boldsymbol{\theta}_1}e^{ir_1^{\prime}\boldsymbol{k}_{\perp}^{\prime}\cdot(\boldsymbol{\theta}_1+\boldsymbol{R}/r_1)}e^{ir_2\boldsymbol{q}_{\perp}\cdot\boldsymbol{\theta_2}}e^{ir_2^{\prime}\boldsymbol{q}_{\perp}^{\prime}\cdot(\boldsymbol{\theta}_2+\boldsymbol{S}/r_2)}\nonumber\\
&\times e^{ir_1k_{\shortparallel}}e^{ir_1^{\prime}k_{\shortparallel}^{\prime}}e^{ir_2q_{\shortparallel}}e^{ir_2^{\prime}q_{\shortparallel}^{\prime}},\\
 =  &\;\frac{1}{(2\pi)^6}\int\int d^3kd^3k^{\prime}\:P(k)P(k^{\prime})e^{ir_1\boldsymbol{k}_{\perp}\cdot\boldsymbol{\theta}_1}e^{ir_1^{\prime}\boldsymbol{k}_{\perp}^{\prime}\cdot(\boldsymbol{\theta}_1+\boldsymbol{R}/r_1)}e^{-ir_2\boldsymbol{k}_{\perp}\cdot\boldsymbol{\theta_2}}e^{-ir_2^{\prime}\boldsymbol{k}_{\perp}^{\prime}\cdot(\boldsymbol{\theta}_2+\boldsymbol{S}/r_2)} e^{ik_{\shortparallel}(r_1-r_2)}e^{ik_{\shortparallel}^{\prime}(r_1^{\prime}-r_2^{\prime})}.
\end{align}
We can again separate the integrals:
\begin{align}
A =  &\;\frac{1}{(2\pi)^6}\int\int d^2k_{\perp}d^2k^{\prime}_{\perp}\:P(k_{\perp})P(k^{\prime}_{\perp})e^{ir_1\boldsymbol{k}_{\perp}\cdot\boldsymbol{\theta}_1}e^{ir_1^{\prime}\boldsymbol{k}_{\perp}^{\prime}\cdot(\boldsymbol{\theta}_1+\boldsymbol{R}/r_1)}e^{-ir_2\boldsymbol{k}_{\perp}\cdot\boldsymbol{\theta_2}}e^{-ir_2^{\prime}\boldsymbol{k}_{\perp}^{\prime}\cdot(\boldsymbol{\theta}_2+\boldsymbol{S}/r_2)}\int dk_{\shortparallel}\:e^{ik_{\shortparallel}(r_1-r_2)} \int dk_{\shortparallel}^{\prime}\: e^{ik_{\shortparallel}^{\prime}(r_1^{\prime}-r_2^{\prime})},\\
=&\;\frac{1}{(2\pi)^4}\int\int d^2k_{\perp}d^2k^{\prime}_{\perp}\:P(k_{\perp})P(k^{\prime}_{\perp})e^{ir_1\boldsymbol{k}_{\perp}\cdot\boldsymbol{\theta}_1}e^{ir_1^{\prime}\boldsymbol{k}_{\perp}^{\prime}\cdot(\boldsymbol{\theta}_1+\boldsymbol{R}/r_1)}e^{-ir_2\boldsymbol{k}_{\perp}\cdot\boldsymbol{\theta_2}}e^{-ir_2^{\prime}\boldsymbol{k}_{\perp}^{\prime}\cdot(\boldsymbol{\theta}_2+\boldsymbol{S}/r_2)}\delta_D(r_1-r_2)\delta_D(r_1^{\prime}-r_2^{\prime}).
\end{align}
We may now insert this expression into Equation (\ref{e:cov_num}) to take advantage of the Dirac delta functions:
\begin{align}
\langle A\rangle =&\;b_s^2b_p^2\int dr_1\:n_s(r_1)\mathcal{W}(r_1,\eta)\int dr_1^{\prime}\:r_1^{\prime2}n_p(r_1^{\prime})\int dr_2\:  n_s(r_2)\mathcal{W}(r_2,\eta)\int dr_2^{\prime}\: r_2^{\prime2}n_p(r_2^{\prime})\frac{1}{(2\pi)^4}\int d^2\theta_1\int d^2\theta_2\nonumber\\
&\times \int\int d^2k_{\perp}d^2k^{\prime}_{\perp}\:P(k_{\perp})P(k^{\prime}_{\perp})e^{ir_1\boldsymbol{k}_{\perp}\cdot\boldsymbol{\theta}_1}e^{ir_1^{\prime}\boldsymbol{k}_{\perp}^{\prime}\cdot(\boldsymbol{\theta}_1+\boldsymbol{R}/r_1)}e^{-ir_2\boldsymbol{k}_{\perp}\cdot\boldsymbol{\theta_2}}e^{-ir_2^{\prime}\boldsymbol{k}_{\perp}^{\prime}\cdot(\boldsymbol{\theta}_2+\boldsymbol{S}/r_2)}\delta_D(r_1-r_2)\delta_D(r_1^{\prime}-r_2^{\prime}),\\
=&\frac{b_s^2b_p^2}{(2\pi)^4}\int dr_1\: n_s(r_1)^2\mathcal{W}(r_1,\eta)^2\int dr_1^{\prime}\: r_1^{\prime4}n_p(r_1^{\prime})^2\int\int d^2k_{\perp}d^2k^{\prime}_{\perp}\:P(k_{\perp})P(k^{\prime}_{\perp})e^{i(r_1^{\prime}/r_1)\boldsymbol{k}_{\perp}^{\prime}\cdot\boldsymbol{R}}e^{-i(r_1^{\prime}/r_1)\boldsymbol{k}_{\perp}^{\prime}\cdot\boldsymbol{S}}\nonumber\\
&\times \int d^2\theta_2\int d^2\theta_1\: e^{ir_1\boldsymbol{k}_{\perp}\cdot(\boldsymbol{\theta}_1-\boldsymbol{\theta}_2)}\: e^{ir_1^{\prime}\boldsymbol{k}_{\perp}^{\prime}\cdot(\boldsymbol{\theta}_1-\boldsymbol{\theta}_2)}.
\end{align}
To perform the angular integrals, we define the substitution $\boldsymbol{\gamma} = \boldsymbol{\theta}_1-\boldsymbol{\theta}_2$, which yields
\begin{align}
\langle A\rangle= &\frac{b_s^2b_p^2}{(2\pi)^4}\int dr_1\: n_s(r_1)^2\mathcal{W}(r_1,\eta)^2\int dr_1^{\prime}\: r_1^{\prime4}n_p(r_1^{\prime})^2\int\int d^2k_{\perp}d^2k^{\prime}_{\perp}\:P(k_{\perp})P(k^{\prime}_{\perp})e^{i(r_1^{\prime}/r_1)\boldsymbol{k}_{\perp}^{\prime}\cdot\boldsymbol{R}}e^{-i(r_1^{\prime}/r_1)\boldsymbol{k}_{\perp}^{\prime}\cdot\boldsymbol{S}}\nonumber\\
&\times \int d^2\theta_2\int d^2\gamma\:e^{i(r_1\boldsymbol{k}_{\perp}+r_1^{\prime}\boldsymbol{k}_{\perp}^{\prime})\cdot\gamma}.
\end{align}
The $\gamma$ term will integrate to a Dirac delta function, whereas the lone $\theta_2$ integral will yield the survey area, denoted $\mathcal{A}$. Accordingly,
\begin{align}
\langle A\rangle=&\frac{b_s^2b_p^2}{(2\pi)^4}\int dr_1\: n_s(r_1)^2\mathcal{W}(r_1,\eta)^2\int dr_1^{\prime}\: r_1^{\prime4}n_p(r_1^{\prime})^2\int\int d^2k_{\perp}d^2k^{\prime}_{\perp}\:P(k_{\perp})P(k^{\prime}_{\perp})e^{i(r_1^{\prime}/r_1)\boldsymbol{k}_{\perp}^{\prime}\cdot(\boldsymbol{R}-\boldsymbol{S})}\mathcal{A}(2\pi)^2\delta^{(2)}_D(r_1\boldsymbol{k}_{\perp}+r_1^{\prime}\boldsymbol{k}_{\perp}^{\prime}),\\
=&\frac{b_s^2b_p^2}{(2\pi)^2}\mathcal{A}\int dr_1\:n_s(r_1)^2\mathcal{W}(r_1,\eta)^2\int dr_1^{\prime} \:  r_1^{\prime2}n_p(r_1^{\prime})^2\int  d^2k_{\perp} P(k_{\perp})P(r_1k_{\perp}/r_1^{\prime})e^{-i\boldsymbol{k}_{\perp}\cdot(\boldsymbol{R}-\boldsymbol{S})}.
\end{align}
Now, let $\boldsymbol{u} = r_1\boldsymbol{k}_{\perp}$, so that:
\begin{align}
\langle A\rangle =&\;\frac{b_s^2b_p^2}{(2\pi)^2}\mathcal{A}\int dr_1\: n_s(r_1)^2\mathcal{W}(r_1,\eta)^2\int dr_1^{\prime}  \: r_1^{\prime2} n_p(r_1^{\prime})^2\int \:\frac{d^2u}{r_1^2}P\left(\frac{u}{r_1}\right)P\left(\frac{u}{r_1^{\prime}}\right)e^{i\boldsymbol{u}\cdot(\boldsymbol{S}-\boldsymbol{R})/r_1},\\
=&\;\frac{b_s^2b_p^2}{(2\pi)^2}\mathcal{A}\int \frac{dr_1}{r_1^2}\: n_s(r_1)^2\mathcal{W}(r_1,\eta)^2\int dr_1^{\prime} r_1^{\prime2}n_p(r_1^{\prime})^2\int d^2u\:P\left(\frac{u}{r_1}\right)P\left(\frac{u}{r_1^{\prime}}\right)e^{i\boldsymbol{u}\cdot(\boldsymbol{S}-\boldsymbol{R})/r_1}.
\end{align}

We proceed along similar lines for term $B$.
\begin{align}
 B =  &\;\frac{1}{(2\pi)^6}\int\int\int\int d^3kd^3k^{\prime}d^3qd^3q^{\prime}\:\langle \delta_{\boldsymbol{k}}\delta_{\boldsymbol{q}^{\prime}}\rangle\langle \delta_{\boldsymbol{k}^{\prime}}\delta_{\boldsymbol{q}}\rangle e^{ir_1\boldsymbol{k}_{\perp}\cdot\boldsymbol{\theta}_1}e^{ir_1^{\prime}\boldsymbol{k}_{\perp}^{\prime}\cdot(\boldsymbol{\theta}_1+\boldsymbol{R}/r_1)}e^{ir_2\boldsymbol{q}_{\perp}\cdot\boldsymbol{\theta_2}}e^{ir_2^{\prime}\boldsymbol{q}_{\perp}^{\prime}\cdot(\boldsymbol{\theta}_2+\boldsymbol{S}/r_2)}e^{ir_1k_{\shortparallel}}e^{ir_1^{\prime}k_{\shortparallel}^{\prime}}e^{ir_2q_{\shortparallel}}e^{ir_2^{\prime}q_{\shortparallel}^{\prime}},\\
=  &\;\frac{1}{(2\pi)^6}\int\int\int\int d^3kd^3k^{\prime}d^3qd^3q^{\prime}\:\delta^{(3)}_D(\boldsymbol{k}+\boldsymbol{q}^{\prime})P(q^{\prime}) \delta^{(3)}_D(\boldsymbol{k}^{\prime}+\boldsymbol{q})P(q)e^{ir_1\boldsymbol{k}_{\perp}\cdot\boldsymbol{\theta}_1}e^{ir_1^{\prime}\boldsymbol{k}_{\perp}^{\prime}\cdot(\boldsymbol{\theta}_1+\boldsymbol{R}/r_1)}e^{ir_2\boldsymbol{q}_{\perp}\cdot\boldsymbol{\theta_2}}e^{ir_2^{\prime}\boldsymbol{q}_{\perp}^{\prime}\cdot(\boldsymbol{\theta}_2+\boldsymbol{S}/r_2)}\nonumber\\
&\times e^{ir_1k_{\shortparallel}}e^{ir_1^{\prime}k_{\shortparallel}^{\prime}}e^{ir_2q_{\shortparallel}}e^{ir_2^{\prime}q_{\shortparallel}^{\prime}},\\
=&\;\frac{1}{(2\pi)^6}\int\int d^3kd^3k^{\prime}\:P(k)P(k^{\prime})e^{ir_1\boldsymbol{k}_{\perp}\cdot\boldsymbol{\theta}_1}e^{ir_1^{\prime}\boldsymbol{k}_{\perp}^{\prime}\cdot(\boldsymbol{\theta}_1+\boldsymbol{R}/r_1)}e^{-ir_2\boldsymbol{k}_{\perp}^{\prime}\cdot\boldsymbol{\theta_2}}e^{-ir_2^{\prime}\boldsymbol{k}_{\perp}\cdot(\boldsymbol{\theta}_2+\boldsymbol{S}/r_2)}e^{i(r_1-r_2^{\prime})k_{\shortparallel}}e^{i(r_1^{\prime}-r_2)k_{\shortparallel}^{\prime}}.
\end{align}
We split the integrals, assuming small angles:
\begin{align}
 B  = &\;\frac{1}{(2\pi)^6}\int\int d^2k_{\perp}d^2k^{\prime}_{\perp}\:P(k_{\perp})P(k^{\prime}_{\perp})e^{ir_1\boldsymbol{k}_{\perp}\cdot\boldsymbol{\theta}_1}e^{ir_1^{\prime}\boldsymbol{k}_{\perp}^{\prime}\cdot(\boldsymbol{\theta}_1+\boldsymbol{R}/r_1)}e^{-ir_2\boldsymbol{k}_{\perp}^{\prime}\cdot\boldsymbol{\theta_2}}e^{-ir_2^{\prime}\boldsymbol{k}_{\perp}\cdot(\boldsymbol{\theta}_2+\boldsymbol{S}/r_2)}\int dk_{\shortparallel}\:e^{i(r_1-r_2^{\prime})k_{\shortparallel}}\int dk_{\shortparallel}^{\prime}\:e^{i(r_1^{\prime}-r_2)k_{\shortparallel}^{\prime}},\\
 = &\;\frac{1}{(2\pi)^4}\int\int d^2k_{\perp}d^2k^{\prime}_{\perp}\:P(k_{\perp})P(k^{\prime}_{\perp})e^{ir_1\boldsymbol{k}_{\perp}\cdot\boldsymbol{\theta}_1}e^{ir_1^{\prime}\boldsymbol{k}_{\perp}^{\prime}\cdot(\boldsymbol{\theta}_1+\boldsymbol{R}/r_1)}e^{-ir_2\boldsymbol{k}_{\perp}^{\prime}\cdot\boldsymbol{\theta_2}}e^{-ir_2^{\prime}\boldsymbol{k}_{\perp}\cdot(\boldsymbol{\theta}_2+\boldsymbol{S}/r_2)} \delta_D(r_1-r_2^{\prime})\delta_D(r_1^{\prime}-r_2).
\end{align}
Plugging this into Equation (\ref{e:cov_num}) yields
\begin{align}
\langle B \rangle =&\; b_s^2b_p^2\int dr_1\: n_s(r_1)\mathcal{W}(r_1,\eta)\int dr_1^{\prime}\: r_1^{\prime2}n_p(r_1^{\prime})\int dr_2\: n_s(r_2)\mathcal{W}(r_2,\eta)\int dr_2^{\prime}\: r_2^{\prime2}n_p(r_2^{\prime})\frac{1}{(2\pi)^4}\int d^2\theta_1\int d^2\theta_2\nonumber\\
&\times \int\int d^2k_{\perp}d^2k^{\prime}_{\perp}\:P(k_{\perp})P(k^{\prime}_{\perp})e^{ir_1\boldsymbol{k}_{\perp}\cdot\boldsymbol{\theta}_1}e^{ir_1^{\prime}\boldsymbol{k}_{\perp}^{\prime}\cdot(\boldsymbol{\theta}_1+\boldsymbol{R}/r_1)} e^{-ir_2\boldsymbol{k}_{\perp}^{\prime}\cdot\boldsymbol{\theta_2}}e^{-ir_2^{\prime}\boldsymbol{k}_{\perp}\cdot(\boldsymbol{\theta}_2+\boldsymbol{S}/r_2)}\delta_D(r_1-r_2^{\prime})\delta_D(r_1^{\prime}-r_2),\\
 =&\; \frac{b_s^2b_p^2}{(2\pi)^4}\int dr_1\: r_1^2n_s(r_1)\mathcal{W}(r_1,\eta)n_p(r_1)\int dr_1^{\prime}\: r_1^{\prime2}n_s(r_1^{\prime})\mathcal{W}(r_1^{\prime},\eta)n_p(r_1^{\prime}) \int\int d^2k_{\perp}d^2k^{\prime}_{\perp}\:P(k_{\perp})P(k^{\prime}_{\perp})e^{i(r_1^{\prime}/r_1)\boldsymbol{k}_{\perp}^{\prime}\cdot\boldsymbol{R}}e^{-i(r_1/r_1^{\prime})\boldsymbol{k}_{\perp}\cdot\boldsymbol{S}}\nonumber\\
&\times\int d^2\theta_1 \:e^{i(r_1\boldsymbol{k}_{\perp}+r_1^{\prime}\boldsymbol{k}_{\perp}^{\prime})\cdot\boldsymbol{\theta}_1}\int d^2\theta_2\:e^{-i(r_1\boldsymbol{k}_{\perp}+r_1^{\prime}\boldsymbol{k}_{\perp}^{\prime})\cdot\boldsymbol{\theta}_2},\\
=&\; \frac{b_s^2b_p^2}{(2\pi)^4}\int dr_1\: r_1^2n_s(r_1)\mathcal{W}(r_1,\eta)n_p(r_1)\int dr_1^{\prime}\: r_1^{\prime2}n_s(r_1^{\prime})\mathcal{W}(r_1^{\prime},\eta)n_p(r_1^{\prime}) \int\int d^2k_{\perp}d^2k^{\prime}_{\perp}\:P(k_{\perp})P(k^{\prime}_{\perp})e^{i(r_1^{\prime}/r_1)\boldsymbol{k}_{\perp}^{\prime}\cdot\boldsymbol{R}}e^{-i(r_1/r_1^{\prime})\boldsymbol{k}_{\perp}\cdot\boldsymbol{S}}\nonumber\\
&\times \mathcal{A}(2\pi)^2\delta_D^{(2)}(r_1\boldsymbol{k}_{\perp}+r_1^{\prime}\boldsymbol{k}_{\perp}^{\prime}),\\
=& \frac{b_s^2b_p^2}{(2\pi)^2}\mathcal{A}\int dr_1\: r_1^2n_s(r_1)\mathcal{W}(r_1,\eta)n_p(r_1)\int dr_1^{\prime}\: n_s(r_1^{\prime})\mathcal{W}(r_1^{\prime},\eta)n_p(r_1^{\prime}) \int d^2k_{\perp}\:P(k_{\perp})P(r_1k_{\perp}/r_1^{\prime})e^{-i\boldsymbol{k}_{\perp}\cdot\boldsymbol{R}}e^{-i(r_1/r_1^{\prime})\boldsymbol{k}_{\perp}\cdot\boldsymbol{S}}.
\end{align}
We set $\boldsymbol{u} = r_1\boldsymbol{k}_{\perp}$ again, so that
\begin{align}
\langle B \rangle =& \frac{b_s^2b_p^2}{(2\pi)^2}\mathcal{A}\int dr_1\: r_1^2n_s(r_1)\mathcal{W}(r_1,\eta)n_p(r_1)\int dr_1^{\prime}\: n_s(r_1^{\prime})\mathcal{W}(r_1^{\prime},\eta)n_p(r_1^{\prime}) \int d^2u\:\frac{1}{r_1^2}P\left(\frac{u}{r_1}\right)P\left(\frac{u}{r_1^{\prime}}\right)e^{-i\boldsymbol{u}\cdot\boldsymbol{R}/r_1}e^{-i\boldsymbol{u}\cdot\boldsymbol{S}/r_1^{\prime}},\\
=& \frac{b_s^2b_p^2}{(2\pi)^2}\mathcal{A}\int dr_1\: n_s(r_1)\mathcal{W}(r_1,\eta)n_p(r_1)\int dr_1^{\prime}\: n_s(r_1^{\prime})\mathcal{W}(r_1^{\prime},\eta)n_p(r_1^{\prime}) \int d^2u\:P\left(\frac{u}{r_1}\right)P\left(\frac{u}{r_1^{\prime}}\right)e^{-i\boldsymbol{u}\cdot(\boldsymbol{R}/r_1+\boldsymbol{S}/r_1^{\prime})}
\end{align}
Lastly, the denominator of the covariance is given by
\begin{align}
D_C&= \int dr_1\: n_s(r_1)\mathcal{W}(r_1,\eta)\int dr_1^{\prime}\:r_1^{\prime2}n_p(r_1^{\prime})\int dr_2\:  n_s(r_2)\mathcal{W}(r_2,\eta)\int dr_2^{\prime}\: r_2^{\prime2}n_p(r_2^{\prime})\int d^2\theta_1\int d^2\theta_2,\\
& = \mathcal{A}^2\int dr_1\: n_s(r_1)\mathcal{W}(r_1,\eta)\int dr_1^{\prime}\:r_1^{\prime2}n_p(r_1^{\prime})\int dr_2\:  n_s(r_2)\mathcal{W}(r_2,\eta)\int dr_2^{\prime}\:r_2^{\prime2} n_p(r_2^{\prime}).
\end{align}

Accordingly, with $c(\boldsymbol{R},\boldsymbol{S}) = \langle N_C\rangle /D_C$, we obtain
\begin{align}\label{e:cov_orig}
c(\boldsymbol{R},\boldsymbol{S}) =&\; \frac{b_s^2b_p^2}{4\pi^2\mathcal{A}}\biggl(\int \frac{dr_1}{r_1^2}\: n_s(r_1)^2\mathcal{W}(r_1,\eta)^2\int dr_1^{\prime}\: r_1^{\prime2}n_p(r_1^{\prime})^2\int d^2u\:P\left(\frac{u}{r_1}\right)P\left(\frac{u}{r_1^{\prime}}\right)e^{i\boldsymbol{u}\cdot(\boldsymbol{S}-\boldsymbol{R})/r_1}\nonumber\\
&+\int dr_1\:  n_s(r_1)\mathcal{W}(r_1,\eta)n_p(r_1) \int dr_1^{\prime}\: n_s(r_1^{\prime})\mathcal{W}(r_1^{\prime},\eta)n_p(r_1^{\prime})\int d^2u\:P\left(\frac{u}{r_1}\right)P\left(\frac{u}{r_1^{\prime}}\right)e^{-i\boldsymbol{u}\cdot(\boldsymbol{R}/r_1+\boldsymbol{S}/r_1^{\prime})}\biggr)\nonumber\\
&\times\left[\int\! dr_1 n_s(r_1)\mathcal{W}(r_1,\eta)\!\int \!dr_1^{\prime}r_1^{\prime2}n_p(r_1^{\prime})\!\int\! dr_2 n_s(r_2)\mathcal{W}(r_2,\eta)\!\int \!dr_2^{\prime} r_2^{\prime2}n_p(r_2^{\prime})\right]^{-1}.
\end{align}
We see that the covariance matrix has two distinct terms: the first integrates over the spectroscopic and photometric samples separately, whereas the second mixes the two samples. 

\subsection{Binned Correlation Function}
While so far we have focused on a single separation $\bf R$, in most cosmological theories, the clustering is independent of the angle on the sky and therefore it is usual for one to integrate over circular annuli, whose bounds we denote by $R_1$ and $R_2$, to produce a binned correlation function as follows:
\begin{align}
w(R_1,R_2) = 2\int_{R_1}^{R_2}\frac{R\:dR}{\left(R_2^2-R_1^2\right)}\int\frac{d\phi}{2\pi}\langle W(\boldsymbol{R})\rangle.
\end{align}
Inserting the form in Equation~\ref{e:final_w} gives
\begin{align}
\langle W(\boldsymbol{R}) \rangle &= \frac{b_sb_p}{2\pi}w_p(\boldsymbol{R})\frac{\int dr \:r^2n_s(r)\mathcal{W}(r,\eta)n_p(r) }{\int dr  \: n_s(r)\mathcal{W}(r,\eta)\int dr^{\prime}  \:r^{\prime2}n_p(r^{\prime})} = \frac{b_sb_p}{(2\pi)^2} \int d^2k_{\perp}^{\prime}\: P(k_{\perp}^{\prime}) e^{i\boldsymbol{k}_{\perp}^{\prime}\cdot\boldsymbol{R}}\frac{\int dr \:r^2n_s(r)\mathcal{W}(r,\eta)n_p(r) }{\int dr  \: n_s(r)\mathcal{W}(r,\eta)\int dr^{\prime}  \:r^{\prime2}n_p(r^{\prime})}, 
\end{align}
and writing $\boldsymbol{k_{\perp}}^{\prime}\cdot\boldsymbol{R} = k_{\perp}^{\prime}R\cos(\phi)$, we obtain the following representation in terms of the Bessel function $J_0$:
\begin{align}
w(R_1,R_2) &= \frac{b_sb_p}{4\pi^3\left(R_2^2-R_1^2\right)}\frac{\int dr \:r^2n_s(r)\mathcal{W}(r,\eta)n_p(r) }{\int dr  \: n_s(r)\mathcal{W}(r,\eta)\int dr^{\prime}  \:r^{\prime2}n_p(r^{\prime})}\int d^2k_{\perp}^{\prime}\: P(k_{\perp}^{\prime})\int_{R_1}^{R_2}R\: dR\int d\phi \:e^{ik_{\perp}^{\prime}R\cos(\phi)},\\
w(R_1,R_2) &= \frac{b_sb_p}{4\pi^3\left(R_2^2-R_1^2\right)}\frac{\int dr \:r^2n_s(r)\mathcal{W}(r,\eta)n_p(r) }{\int dr  \: n_s(r)\mathcal{W}(r,\eta)\int dr^{\prime}  \:r^{\prime2}n_p(r^{\prime})}\int d^2k_{\perp}^{\prime}\: P(k_{\perp}^{\prime})\int_{R_1}^{R_2}R\: dR\: 2\pi J_0(k_{\perp}^{\prime}R).
\end{align}
The integral over $R$ can now be done by making use of the properties of Bessel functions, leading to a final result of
\begin{align}\label{e:binw}
w(R_1,R_2) &= \frac{b_sb_p}{2\pi^2\left(R_2^2-R_1^2\right)}\frac{\int dr \:r^2n_s(r)\mathcal{W}(r,\eta)n_p(r) }{\int dr  \: n_s(r)\mathcal{W}(r,\eta)\int dr^{\prime}  \:r^{\prime2}n_p(r^{\prime})}\int d^2k_{\perp}^{\prime}\: P(k_{\perp}^{\prime})\left[\frac{R_2J_1(k_{\perp}^{\prime}R_2)-R_1J_1(k_{\perp}^{\prime}R_1)}{k_{\perp}^{\prime}}\right].
\end{align}

\subsection{Binned Covariance Matrix}\label{s:bincov}
We also compute the binned covariance matrix. In this case we have two terms to consider (temporarily omitting constant values to focus on the derivation):
\begin{align}
c_1 &\sim \int d^2u\:P\left(\frac{u}{r_1}\right)P\left(\frac{u}{r_1^{\prime}}\right)e^{i\boldsymbol{u}\cdot(\boldsymbol{S}-\boldsymbol{R})/r_1},\\
c_2 &\sim \int d^2u\:P\left(\frac{u}{r_1}\right)P\left(\frac{u}{r_1^{\prime}}\right)e^{-i\boldsymbol{u}\cdot(\boldsymbol{R}/r_1+\boldsymbol{S}/r_1^{\prime})}.
\end{align}
Accordingly, for the first term we find
\begin{align}
C_{1} &\sim 4\int_{R_1}^{R_2}\frac{R\:dR}{\left(R_2^2-R_1^2\right)}\int\frac{d\phi_R}{2\pi}\int_{S_1}^{S_2}\frac{S\:dS}{\left(S_2^2-S_1^2\right)}\int\frac{d\phi_S}{2\pi}\int d^2u\:P\left(\frac{u}{r_1}\right)P\left(\frac{u}{r_1^{\prime}}\right)e^{i\boldsymbol{u}\cdot(\boldsymbol{S}-\boldsymbol{R})/r_1},
\end{align}
which, with $\boldsymbol{u}\cdot\boldsymbol{R} = uR\cos(\phi_R)$ and $\boldsymbol{u}\cdot\boldsymbol{S} = uS\cos(\phi_S)$, becomes
\begin{align}
C_{1} \sim& \frac{1}{\pi^2\left(R_2^2-R_1^2\right)\left(S_2^2-S_1^2\right)}\int d^2u\:P\left(\frac{u}{r_1}\right)P\left(\frac{u}{r_1^{\prime}}\right)\int R\:dR\int d\phi_R\:e^{iuR\cos(\phi_R)/r_1}\int S\:dS\int d\phi_S\:e^{-iuS\cos(\phi_S)/r_1},\\
\sim& \frac{4}{\left(R_2^2-R_1^2\right)\left(S_2^2-S_1^2\right)}\int \! d^2u\:P\left(\frac{u}{r_1}\right)\! P\left(\frac{u}{r_1^{\prime}}\right)\! \!\left[\frac{r_1^2}{u^2}\right]\! \!\left[R_2J_1\left(\frac{uR_2}{r_1}\right)\!\! -\! R_1J_1\left(\frac{uR_1}{r_1}\right)\right]\!\!\left[S_2J_1\left(\frac{uS_2}{r_1}\right)\!\!-\!S_1J_1\left(\frac{uS_1}{r_1}\right)\right]\!\!.
\end{align}
For the second term, 
\begin{align}
C_{2} \sim & 4\int_{R_1}^{R_2}\frac{R\:dR}{\left(R_2^2-R_1^2\right)}\int\frac{d\phi_R}{2\pi}\int_{S_1}^{S_2}\frac{S\:dS}{\left(S_2^2-S_1^2\right)}\int\frac{d\phi_S}{2\pi}\int d^2u\:P\left(\frac{u}{r_1}\right)P\left(\frac{u}{r_1^{\prime}}\right)e^{-i\boldsymbol{u}\cdot(\boldsymbol{R}/r_1+\boldsymbol{S}/r_1^{\prime})},\\
\sim& \frac{1}{\pi^2\left(R_2^2-R_1^2\right)\left(S_2^2-S_1^2\right)}\int d^2u\:P\left(\frac{u}{r_1}\right)P\left(\frac{u}{r_1^{\prime}}\right)\int R\:dR\int d\phi_R\:e^{-iuR\cos(\phi_R)/r_1}\int S\:dS\int d\phi_S\:e^{-iuS\cos(\phi_S)/r_1^{\prime}},\\
\sim & \frac{4}{\left(R_2^2-R_1^2\right)\left(S_2^2-S_1^2\right)}\int d^2u\:P\left(\frac{u}{r_1}\right)\!P\left(\frac{u}{r_1^{\prime}}\right)\!\!\left[\frac{r_1r_1^{\prime}}{u^2}\right]\!\!\left[R_2J_1\left(\frac{uR_2}{r_1}\right)\!\!-\!R_1J_1\left(\frac{uR_1}{r_1}\right)\right]\!\!\left[S_2J_1\left(\frac{uS_2}{r_1^{\prime}}\right)\!\!-\!S_1J_1\left(\frac{uS_1}{r_1^{\prime}}\right)\right]\!\!.
 \end{align}
 Thus, the binned covariance becomes (c.f. Eq.~\ref{e:cov_orig})
 \begin{align}\label{e:bincov}
C(R_1,R_2,S_1,S_2) =& \frac{b_s^2b_p^2\left[\int dr_1\: n_s(r_1)\mathcal{W}(r_1,\eta)\int dr_1^{\prime}\:r_1^{\prime2}n_p(r_1^{\prime})\int dr_2\:  n_s(r_2)\mathcal{W}(r_2,\eta)\int dr_2^{\prime}\: r_2^{\prime2}n_p(r_2^{\prime})\right]^{-1}}{\pi^2\mathcal{A}\left(R_2^2-R_1^2\right)\left(S_2^2-S_1^2\right)}\nonumber\\
& \times\biggl(\int dr_1 \:n_s(r_1)^2\mathcal{W}(r_1,\eta)^2\int dr_1^{\prime}\: r_1^{\prime2}n_p(r_1^{\prime})^2\int \frac{d^2u}{u^2}\:P\left(\frac{u}{r_1}\right)P\left(\frac{u}{r_1^{\prime}}\right)\nonumber\\
& \;\;\;\;\;\;\times\left[R_2J_1\left(\frac{uR_2}{r_1}\right)-R_1J_1\left(\frac{uR_1}{r_1}\right)\right] \left[S_2J_1\left(\frac{uS_2}{r_1}\right)-S_1J_1\left(\frac{uS_1}{r_1}\right)\right]\nonumber\\
&+\! \int \! dr_1\:  r_1 n_s(r_1)\mathcal{W}(r_1,\eta)n_p(r_1) \! \int\!  dr_1^{\prime}\: r_1^{\prime}n_s(r_1^{\prime})\mathcal{W}(r_1^{\prime},\eta)n_p(r_1^{\prime})\int \frac{d^2u}{u^2}\:P\left(\frac{u}{r_1}\right)\! P\left(\frac{u}{r_1^{\prime}}\right)\nonumber\\
&\times\left[R_2J_1\left(\frac{uR_2}{r_1}\right)-R_1J_1\left(\frac{uR_1}{r_1}\right)\right]\left[S_2J_1\left(\frac{uS_2}{r_1^{\prime}}\right)-S_1J_1\left(\frac{uS_1}{r_1^{\prime}}\right)\right]\biggr).
 \end{align}
 
This expression can be numerically integrated up to the factors of $b_s$ and $b_p$, which may not be known for the datasets under consideration. A form for $n_p(r)$ needs to be assumed as well, and can be estimated from smaller areas with existing spectroscopic catalogs or high quality photometric redshifts. Care must be taken when integrating the Bessel functions, as in the first term, both sets depend on $r_1$, whereas the second term also includes $r_1^{\prime}$; the integration over $u$ thus needs to account for the range of arguments that will be encountered in order to appropriately sample the oscillations. 
 
 \section{Data}
In order to test this method, we use two samples from the Sloan Digital Sky Survey. First, for our spectroscopic dataset, we consider the color cut-defined CMASS sample of galaxies from Data Release 12 of SDSS-III \citep{eisenstein11,dawson13,sdssdr12,reid16} for our spectroscopic sample. These data were obtained with a 2.5~m telescope at Apache Point Observatory \citep{york00,gunn06} whose instrumentation is well-documented \citep{fukugita96,smith02,doi10,smee13} and for which a set of data processing pipelines has been developed \citep{lupton01,pier03,padmanabhan08,bolton12,weaver15}. To obtain a fairly sparse sample of spectra for the validation of the method in the next section, we restrict the redshift range, focusing only on those galaxies for which $z>0.6$. This yields 205,367 galaxies, and we select a set of random points 10 times larger than this for the construction of the correlation function. We note that while this $z>0.6$ data is fairly sparse, it is not a sample for which $\tau<0.08$. The photometric sample used in this paper comes from SDSS DR9 \citep{sdssdr9} imaging data and comprises 6,594,677 galaxies (see \citet{law-smith17} for a detailed description of the color cuts used to define this sample); we use a catalog of randoms five times larger than this to construct the correlation function. 

In practice, the \textit{projected} cross-correlation function between the spectroscopic and photometric samples can be constructed via the Landy-Szalay estimator \citep{landy93},
\begin{align}
w(R) = \frac{D_sD_p(R)-D_sR_p(R)-D_pR_s(R)+R_sR_p(R)}{R_sR_p(R)},
\end{align}
where $D$ indicates the data and $R$ denotes the random points; $p$ and $s$ refer to the photometry and spectroscopy, respectively. While the Landy-Szalay estimator is normally defined as providing an estimate of $\xi(r)$, the full three-dimensional correlation function, in our case we only know the positions of the galaxies on the sky for the photometric sample. Accordingly, when we compute the correlation function, we use only spectroscopic galaxies as our central points and then calculate pair counts in two-dimensional annular bins around the central galaxy, assuming that the photometric galaxies in the bin are at the same redshift as the spectroscopic galaxy to convert angular separations into physical distances, yielding $w(R)$.

We compute errors on our measurements in two ways. First, we divide the spectroscopic survey area into 150 roughly commensurate regions using Voronoi tessellation and then employ the jackknife method. Second, we use the analytical derivation presented in the preceding sections and test the explicit form for the covariance matrix (Equation~\ref{e:bincov}) in fitting the functional form of $w(R)$ given by Equation~\ref{e:binw}.

\section{Results}
\subsection{Fitting for the BAO Peak}
\begin{figure}
\begin{center}
\includegraphics[scale=0.49,trim={0.1cm 0.0cm 0.2cm 0.1cm},clip]{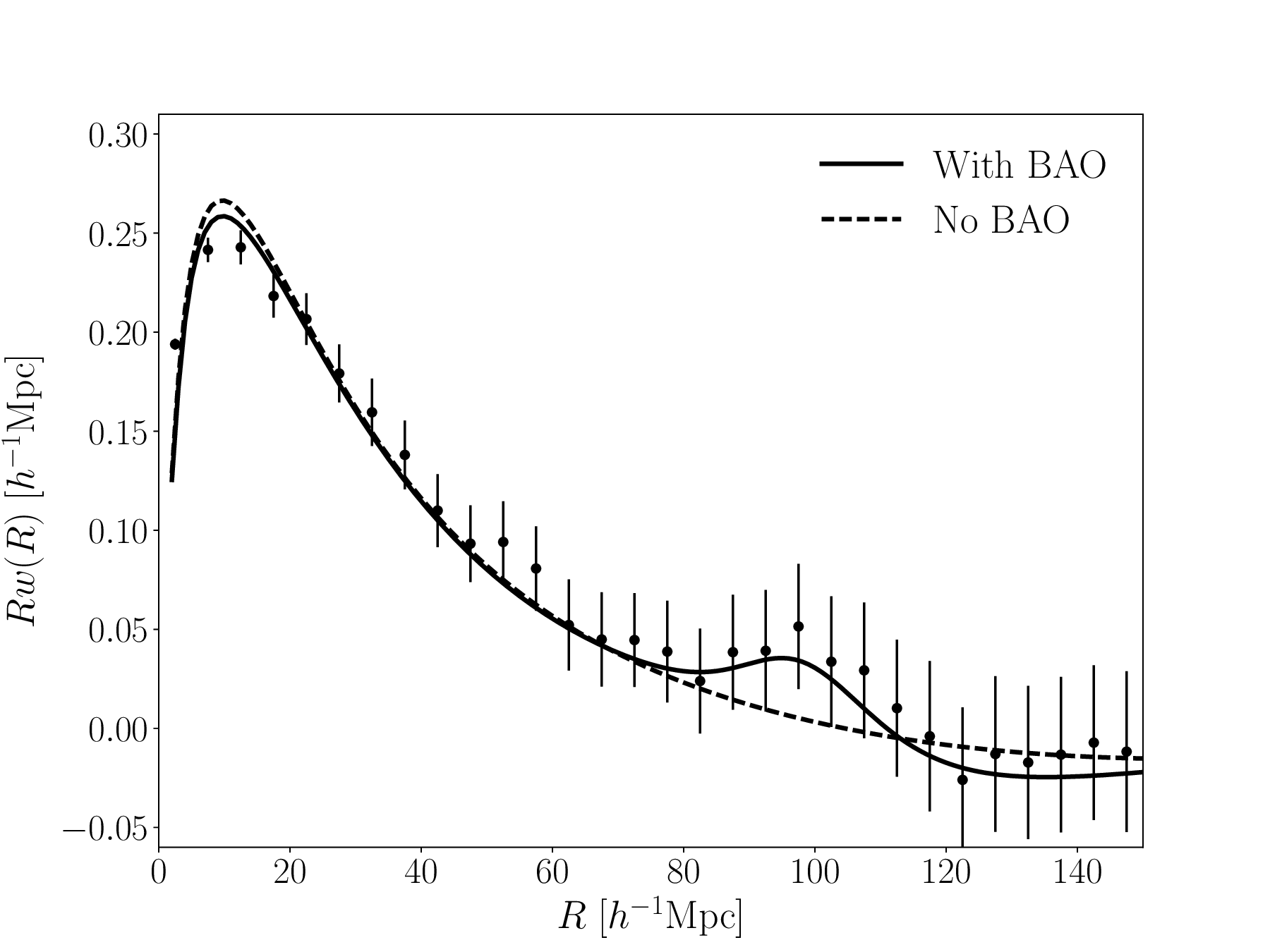}
\end{center}
\caption{The projected correlation function $w$ plotted against two theoretical curves: the solid black line is the predicted projected correlation function in the presence of BAO, while the dashed violet curve is the expectation without BAO. The data show a bump near $100\;h^{-1}\mathrm{Mpc}$ consistent with the BAO signal.}\label{f:wp}
\end{figure}

We compute the projected correlation function $w(R)$ using the Landy-Szalay estimator with a binning of $5\;h^{-1}\mathrm{Mpc}$; the result is plotted in Figure~\ref{f:wp}. We fit two computed projected correlation functions, one derived using a CAMB power spectrum~\citep{camb} that includes the BAO, and the other a smooth, BAO-less power spectrum derived using the prescriptions of \citet{eisenstein98}, to the data. 

\begin{figure}
\begin{center}
\includegraphics[scale=0.35,trim={0.0cm 0.0cm 0.2cm 0.2cm},clip]{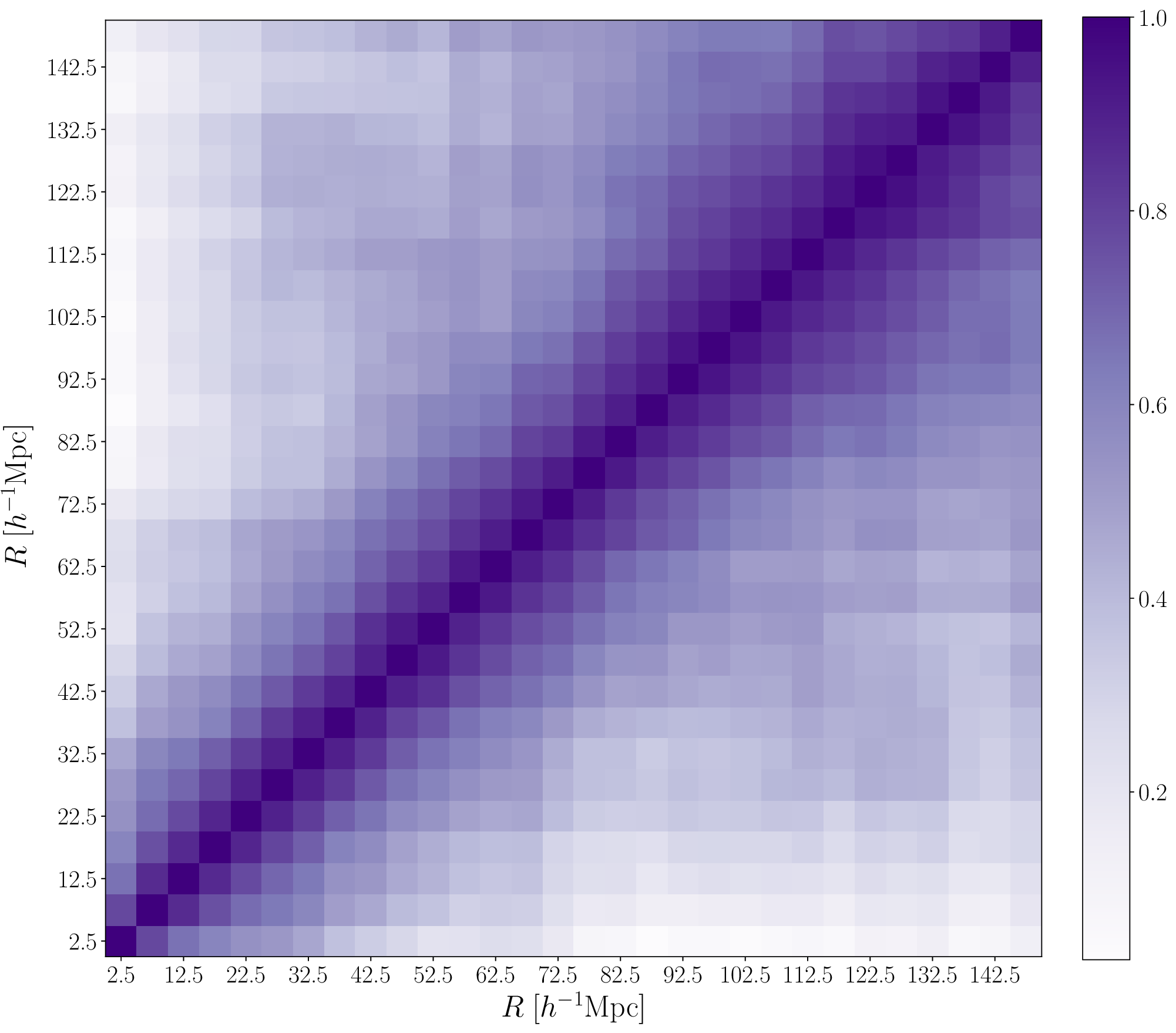}
\end{center}
\caption{The reduced covariance matrix for the cross-correlation, estimated from Voronoi jackknife samples and used to fit theoretical curves to the computed Landy-Szalay $w$.}\label{f:covm}
\end{figure}

In both theoretical prescriptions of the correlation function, we include a nuisance parameter in the form of an additive constant $w_0$, which corresponds to excess power at very large scales and which is left a free parameter. Accordingly, we fit to a function of the form $w(R) = aw_c(R)+w_0$, where $w_c(R)$ is the calculated correlation function and $a$ and $w_0$ are parameters to be found. Prior to fitting, we normalize $w_c$ to the value of the data at $R = 27.5 \;h^{-1}\mathrm{Mpc}$. We use the jackknife covariance matrix shown in Figure~\ref{f:covm} in performing the fits. The results of our fits suggest that the parameter $w_0$ is around $1\times10^{-4}$, indicating only a modest level of systematics.

\begin{figure}
\begin{center}
\includegraphics[scale=0.49,trim={0.0cm 0.0cm 0.2cm 0.3cm},clip]{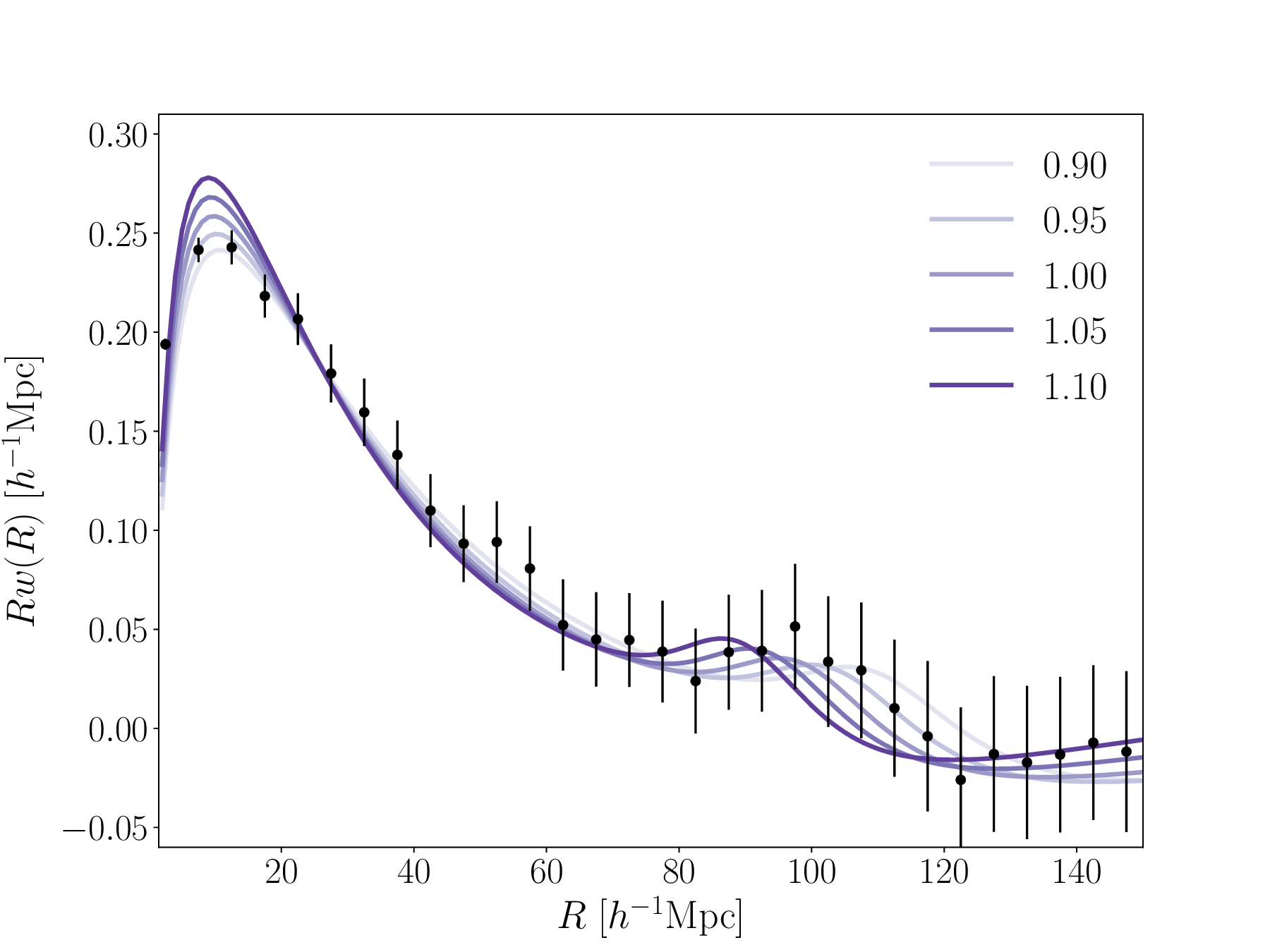}\:\:
\includegraphics[scale=0.49,trim={0.0cm 0.0cm 0.2cm 0.3cm},clip]{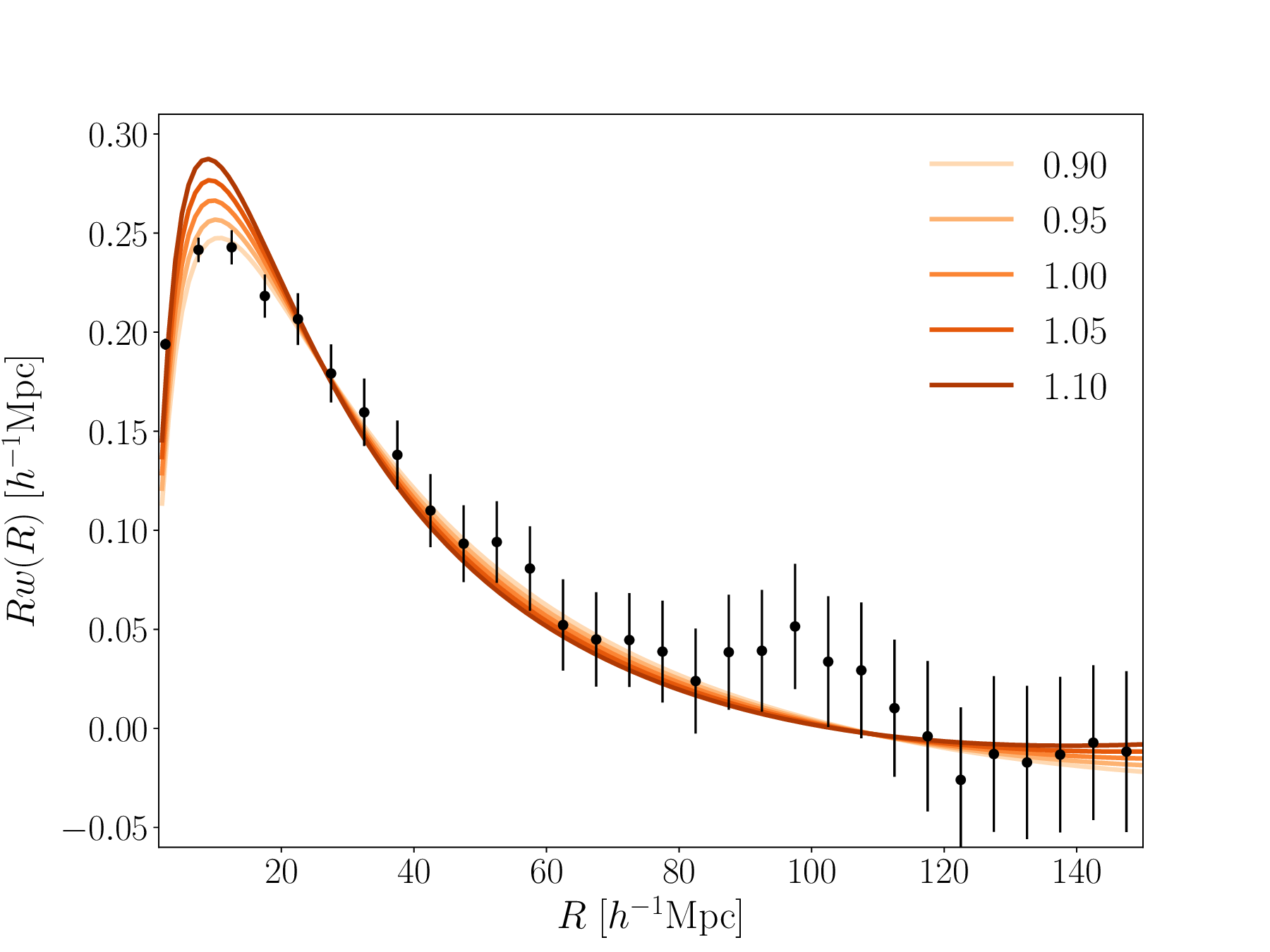}
\end{center}
\caption{\textit{Top:} the projected correlation functions with BAO for various values of the peak shift parameter, $\alpha$, fitted to the data. The data prefer a value of $\alpha=1.00$. \textit{Bottom:} the projected correlation functions for models without the BAO.}\label{f:wpshift}
\end{figure}

\begin{figure}
\begin{center}
\includegraphics[scale=0.5,trim={0.0cm 0.0cm 0.2cm 0.2cm},clip]{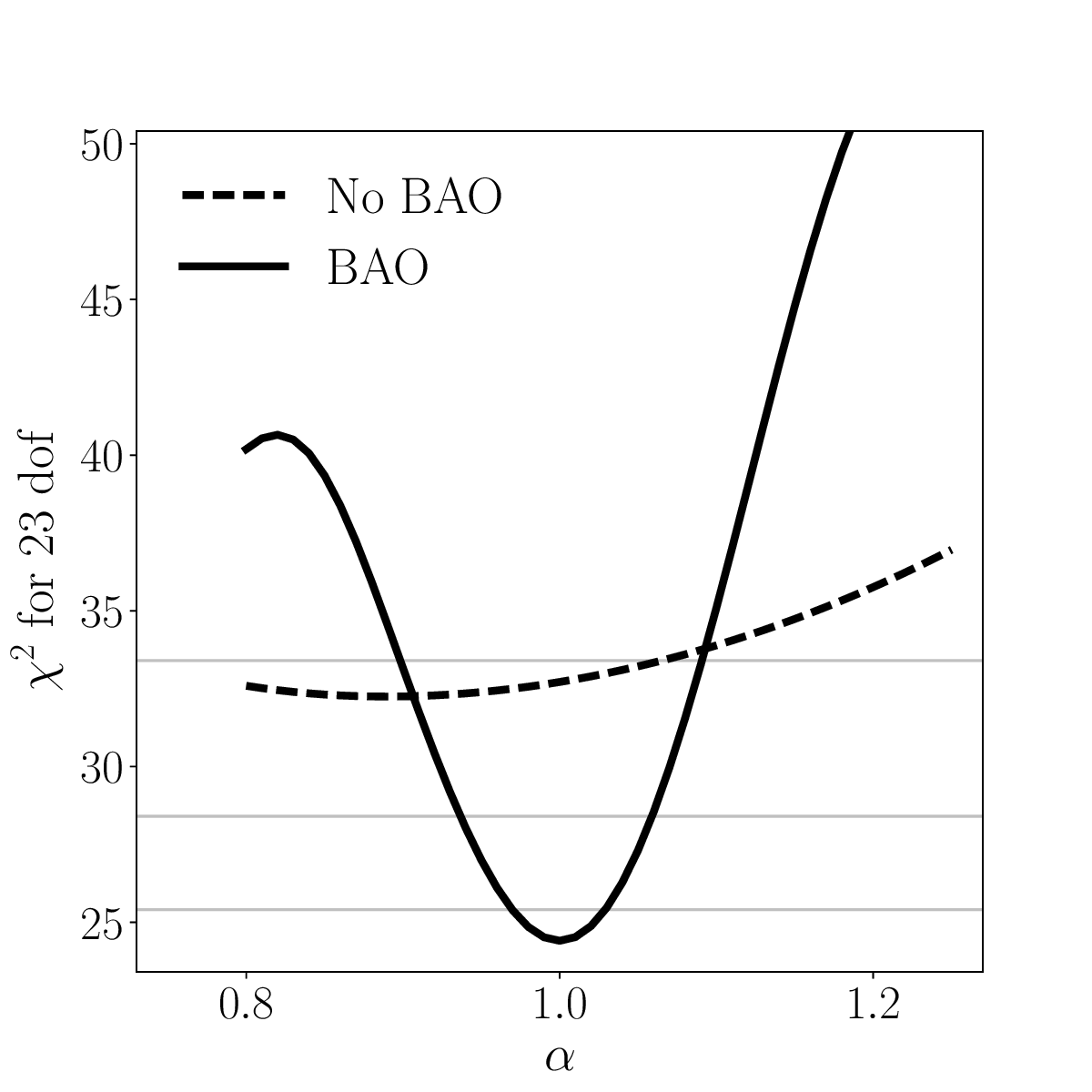}
\end{center}
\caption{The $\chi^2$ values of the fits illustrated in Figure~\ref{f:wpshift}; the gray lines denote the $1\sigma$, $2\sigma$, and $3\sigma$ bounds. The data clearly prefer a model that includes the BAO peak with $\alpha=1.00$.}\label{f:chisq}
\end{figure}

We further consider the impact of a BAO peak shift, quantitatively provided by a dilation factor (denoted $\alpha$) on our results. Scaling $P(k)\rightarrow P(k/\alpha)$ for both the BAO and BAO-less cases, we recompute the correlation functions and fit them to the data, allowing for $\alpha$ to vary in the range of $0.8\lesssim \alpha \lesssim 1.25$; the results for five values of $\alpha$ are shown in Figure~\ref{f:wpshift}. The data prefer a value of $\alpha = 1.00$, as indicated by the $\chi^2$ curves for the fits that are shown in Figure~\ref{f:chisq}. A shift of $\alpha = 1.00$ gives a $\chi^2/\mathrm{ndf} = 24.4/23$, while the best fit value for the non-BAO curve has $\chi^2/\mathrm{ndf} = 32.3/23$, yielding a $2.8\sigma$ preference for the BAO. We do note, however, that as we are using a jackknife method, we have not included corrections for the impact of noise of the inversion of the covariance matrix such as is described in~\citet{percival14} in the context of mock catalogs. 

Interpreting the $\Delta\chi^2 = 4$ range as a $2\sigma$ error, we find a rough estimate of about 0.03 for the $1\sigma$ precision of $\alpha$. However, we caution that the low significance of the acoustic peak detection likely implies that the posterior of $\alpha$ has non-Gaussian tails that would loosen the constraining power. 

Our results align well with the prediction for the correlation function with BAO at intermediate-to-large scales. We note that while we do not have to consider redshift-space distortions with $w$, since we do not account for where the photometric galaxies are located along the line-of-sight, there are non-linear corrections that could have some impact at small scales ($R\lesssim10\;h^{-1}\mathrm{Mpc}$), but these are not expected to yield major changes at large scales. Accordingly, our measurements yield a feature consistent with the BAO signal at $R\sim100\;h^{-1}\mathrm{Mpc}$ in the projected correlation function.

In order to measure the comoving angular diameter distance $D_M(z) = (1+z)D_A(z)$ \citep{padmanabhan08}, where $D_A$ is the physical angular diameter distance, we require an estimate of the effective redshift for our sample, which should be the mean redshift of the spectroscopic and photometric pairs. While we know the spectroscopic redshift distribution, we do not have detailed knowledge of the photometric redshift distribution. If the photometric galaxies had the same distribution as the spectroscopic sample, then the mean pair redshift would be the volume-weighted average of $n_s^2(z)$. However, the photometric sample is 1 magnitude fainter than the spectroscopic sample and has a slightly redder color selection, so we expect the distribution of these galaxies to extend to higher redshift. Alternatively, if the photometric galaxies had a uniform $n_p(z)$, then the mean pair redshift would be the volume-weighted average of $n_s(z)$. Accordingly, we adopt the average of the mean redshifts of $n_s(z)$ and $n_s^2(z)$, $z = 0.64$, as the effective redshift for our sample. 

From the dilation parameter $\alpha$ we then obtain $D_M(z)$ as \citep[see e.g.,][]{alam17}:
\begin{align}
\alpha = \frac{D_M(z)r_{s,\mathrm{fid}}}{D_M^{\mathrm{fid}}r_s},
\end{align}
where $r_s$ denotes the sound horizon and the superscript `fid' refers to our fiducial cosmology. With the results above, we find $D_M(z = 0.64) = (2418 \pm 73\;\mathrm{Mpc}) \left(r_s/r_{s,\mathrm{fid}}\right)$. As a comparison, \cite{alam17} find $D_M(z=0.57) = (2179\pm35\;\mathrm{Mpc})\left(r_s/r_{s,\mathrm{fid}}\right)$. We note that our result is at least partially covariant with their high redshift sample.

\subsection{Comparison to the Theoretical Covariance Matrix}
In obtaining the above results, we have relied on the covariance matrix that we calculated from jackknife resampling. We now return to the analytical expression for the covariance matrix derived in  Section~\ref{s:bincov}, albeit in an idealized form with some unknown quantities. In particular, in our derivation, we neglected to include shot noise and our matrix is only known up to the factor of $b_sb_p$, the product of the spectroscopic and photometric bias parameters, which is not known for our sample. 

We accounted for these unknown factors in our expression as follows: first, to determine the rough value of $b_sb_p$, we took the ratio of the diagonals of the covariance matrix computed from Equation~\ref{e:bincov} and those of the jackknife matrix. This comparison yielded roughly a factor of 2 in the ratio between large and small scales, indicating that the two matrices are not perfectly matched. The sense of the difference is that the jackknife matrix has larger diagonals at larger scales, which might be the result of additional systematic errors at large angles. We therefore chose to normalize the theory matrix at smaller scales ($\lesssim 60\;h^{-1}\mathrm{Mpc}$) where we expect the result to be less affected by large-scale systematics. 

Second, we established the constant value of the shot noise $P_{\mathrm{shot}}$ to use by matching the eigenvalues of both matrices under the transformation $P(k) \rightarrow P(k)+P_{\mathrm{shot}}$ in Equation~\ref{e:bincov}. By choosing a modest value of $P_{\mathrm{shot}}$, we could bring the eigenvalues into agreement without yielding noticeable deviations in the diagonal ratio. Following this procedure, we found $P_{\mathrm{shot}} = 1500 \;(h^{-1}\mathrm{Mpc})^3$, a value that is commensurate with the level of shot noise in the BOSS survey. With this assumption, we found good agreement between the theoretical and jackknife matrices. Upon fitting the data, we found a preference for the BAO peak with $\alpha=0.99$ ($\chi^2 = 15.7$ for 23 degrees of freedom) with $\Delta\chi^2 = 2.9$. The $\chi^2$ suggests we may be slightly over-normalizing the theoretical matrix.

Furthermore, as noted above, we do see indications of the presence of large-angle observational systematic errors; accordingly, we have opted for relying on the results obtained by using the jackknife matrix. While it is beyond the scope of this work, future applications will require careful analyses of the survey systematics, which will need to be accounted for in the analytical calculation of the covariance matrix.

 \section{Summary and Conclusions}
In this work, we have laid out a detailed formalism for a method of detecting the BAO in sparse spectroscopic datasets by cross-correlating the spectroscopy with photometry, for which greater densities of galaxies can be attained at far smaller cost. We analytically computed the projected correlation function and presented an explicit form for the covariance matrix, which can be calculated numerically and used to fit models to the data. We then applied the method to high-redshift data from SDSS.

We find that this method can reveal the BAO signal in sparse data; in our test data set, which relied on a spectroscopic sample consisting of the $z > 0.6$ tail of the CMASS galaxy distribution cross-correlated with SDSS DR9 photometry, we found a $2.8\sigma$ preference for the BAO and obtained a distance measurement of $D_M(z = 0.64) = (2418 \pm 73\;\mathrm{Mpc}) \left(r_s/r_{s,\mathrm{fid}}\right)$. This implies that cross-correlations of spectroscopy and photometry will provide a viable means with which to extend BAO measurements to higher redshifts even before the completion of major future surveys.

Of particular interest may be the application of this method to detect the BAO in the clustering of high redshift quasars observed by present and upcoming surveys like the extended Baryon Oscillation Spectroscopic Survey (eBOSS) and the Dark Energy Spectroscopic Instrument (DESI). The eBOSS quasar sample, for example, is expected to have an average $nP = 0.040$ \citep{zhao16} over the redshift range $0.6<z<2.2$, which would make this method an effective way to detect the BAO in that sample. These data can be combined with photometry from the DECam Legacy Survey (DECaLS), the Mosaic z-band Legacy Survey (MzLS), and the Beijing-Arizona Sky Survey (BASS), which are currently obtaining deep, high-quality imaging of the SDSS survey area in order to provide target selection for DESI. Additionally, this method could be employed to measure the transverse BAO scale at $z\sim2$ using OIII emission line galaxies that will be discovered by \textit{Euclid} \citep[e.g.,][]{mehta15}.

 \section{Acknowledgments}
We would like to thank Aaron Bray for providing us with the photometric data files used in this paper and for generating the corresponding random catalog. We additionally thank Lehman Garrison for generating the CAMB power spectrum used in our calculations. 
 
A.P. acknowledges support from the NASA Einstein Fellowship program under Grant No. PF6-170157. D.J.E. was supported as a Simons Foundation Investigator. This work was also supported by U.S. Department of Energy Grant No. DE-SC0013718, and made use of tools from the Python packages NumPy \citep{numpy11}, SciPy \citep{oliphant07}, and Matplotlib \citep{hunter07}, and also relied on ROOT \citep[][see also http://root.cern.ch/]{root}.

This paper relies on data from SDSS-III. Funding for SDSS-III has been provided by the Alfred P. Sloan Foundation, the Participating Institutions, the National Science Foundation, and the U.S. Department of Energy Office of Science. The SDSS-III web site is http://www.sdss3.org/.

SDSS-III is managed by the Astrophysical Research Consortium for the Participating Institutions of the SDSS-III Collaboration including the University of Arizona, the Brazilian Participation Group, Brookhaven National Laboratory, Carnegie Mellon University, University of Florida, the French Participation Group, the German Participation Group, Harvard University, the Instituto de Astrofisica de Canarias, the Michigan State/Notre Dame/JINA Participation Group, Johns Hopkins University, Lawrence Berkeley National Laboratory, Max Planck Institute for Astrophysics, Max Planck Institute for Extraterrestrial Physics, New Mexico State University, New York University, Ohio State University, Pennsylvania State University, University of Portsmouth, Princeton University, the Spanish Participation Group, University of Tokyo, University of Utah, Vanderbilt University, University of Virginia, University of Washington, and Yale University.

\bsp	
\label{lastpage}


\begin{thebibliography}{}
\bibitem[Ahn et al.(2012)]{sdssdr9}
Ahn, C. P., et al. 2012, ApJS, 203, 21
\bibitem[Alam et al.(2015)]{sdssdr12}
Alam, S., et al. 2015, ApJS, 219, 12
\bibitem[Alam et al.(2017)]{alam17}
Alam, S., et al. 2017, MNRAS, 470, 2617
\bibitem[Anderson et al.(2012)]{anderson12}
Anderson, L., et al. 2012, MNRAS, 427, 3435
\bibitem[Anderson et al.(2014)]{anderson14}
Anderson, L., et al. 2014, MNRAS, 441, 24
\bibitem[Ata et al.(2017)]{ata17}
Ata, M., et al. 2017, arXiv:1705.06373
\bibitem[Bolton et al.(2012)]{bolton12}
Bolton, A. S., et al. 2012, AJ, 144, 144
\bibitem[Brun \& Rademakers(1997)]{root}
Brun, R., \& Rademakers, F. 1997, Nucl. Inst. \& Meth. in Phys. Res. A, 389, 81 
\bibitem[Coil et al.(2008)]{coil08}
Coil, A. L., et al. 2008, ApJ, 672, 153
\bibitem[Cole et al.(2005)]{cole05}
Cole, S., et al. 2005, MNRAS, 362, 505
\bibitem[Colless et al.(2001)]{colless01}
Colless, M., et al. 2001, MNRAS, 328, 1039
\bibitem[Davis \& Peebles(1983)]{davis83}
Davis, M., \& Peebles, P. J. E., 1983, ApJ, 267, 465 
\bibitem[Dawson et al.(2013)]{dawson13}
Dawson, K. S., et al. 2013, AJ, 145, 10
\bibitem[Dawson et al.(2016)]{dawson16}
Dawson, K. S., et al. 2016, AJ, 151, 44
\bibitem[Dekel \& Lahav(1999)]{dekel99}
Dekel, A., \& Lahav, O. 1999, ApJ, 520, 24
\bibitem[Doi et al.(2010)]{doi10}
Doi, M., et al. 2010, AJ, 139, 1628
\bibitem[Eisenstein \& Hu(1998)]{eisenstein98}
Eisenstein, D. J., \& Hu, W. 1998, ApJ, 496, 605
\bibitem[Eisenstein et al.(2005)]{eisenstein05}
Eisenstein, D. J., et al. 2005, ApJ, 633, 560
\bibitem[Eisenstein et al.(2011)]{eisenstein11}
Eisenstein, D. J., et al. 2011, AJ, 142, 72
\bibitem[Fukugita et al.(1996)]{fukugita96}
Fukugita, M. et al. 1996, AJ, 111, 1748
\bibitem[Gunn et al.(2006)]{gunn06}
Gunn, J. E., et al. 2006, AJ, 131, 2332
\bibitem[Hunter(2007)]{hunter07}
Hunter, J. D. 2007, Computing in Science \& Engineering, 9, 90
\bibitem[Landy \& Szalay(1993)]{landy93}
Landy, S. D., \& Szalay, A. S. 1993, ApJ, 412, 64
\bibitem[Law-Smith \& Eisenstein(2017)]{law-smith17}
Law-Smith, J., \& Eisenstein, D. J. 2017, ApJ, 836, 87
\bibitem[Levi et al.(2013)]{levi13}
Levi, M., et al. (DESI Collaboration), 2013, arXiv: 1308:0847
\bibitem[Lewis, Challinor, \& Lasenby(2000)]{camb}
Lewis, A., Challinor, A., \& Lasenby, A. 2000, ApJ, 538, 473
\bibitem[Lupton et al.(2001)]{lupton01}
Lupton, R., Gunn, J. E., Ivezic, Z., Knapp, G.,  \& Kent, S. 2001, Astronomical Data Analysis Software and Systems X, v.238, 269
\bibitem[Mehta et al.(2015)]{mehta15}
Mehta, V., et al. 2015, ApJ, 811, 141
\bibitem[Nishizawa et al.(2013)]{nishizawa13}
Nishizawa, A. J., Oguri, M., \& Takada, M. 2013, MNRAS, 433, 730
\bibitem[Nuza et al.(2013)]{nuza13}
Nuza, S. E., et al. 2013, MNRAS, 432, 743
\bibitem[Oliphant(2007)]{oliphant07}
Oliphant, T. E. 2007, Computing in Science \& Engineering, 9, 10
\bibitem[Padmanabhan et al.(2008)]{padmanabhan08}
Padmanabhan, N., et al. 2008, ApJ, 674, 1217
\bibitem[Percival et al.(2014)]{percival14}
Percival, W., et al. 2014, MNRAS, 439, 2531
\bibitem[Pier et al.(2003)]{pier03}
Pier, J. R., et al. 2003, AJ, 125, 1559
\bibitem[Reid et al.(2016)]{reid16}
Reid, B., et al. 2016, MNRAS, 455, 1553
\bibitem[Smee et al.(2013)]{smee13}
Smee, S. A., et al. 2013, AJ, 146, 32
\bibitem[Smith et al.(2002)]{smith02}
Smith, J. A., et al. 2002, AJ, 123, 2121
\bibitem[van der Walt, Colbert, \& Varoquaux(2011)]{numpy11}
van der Walt, S., Colbert, S. C., \& Varoquaux, G. 2011, Computing in Science \& Engineering, 13, 22
\bibitem[Weaver et al.(2015)]{weaver15}
Weaver, B., et al. 2015, PASP, 127, 397
\bibitem[Weinberg et al.(2013)]{weinberg13}
Weinberg, D. H., et al. 2013, PhR, 530, 87
\bibitem[York et al.(2000)]{york00}
York, D. G., et al. 2000, AJ, 120, 1579
\bibitem[Zehavi et al.(2011)]{zehavi11}
Zehavi, I., et al. 2011, ApJ, 736, 59
\bibitem[Zhao et al.(2016)]{zhao16}
Zhao, G-B., et al. 2016, MNRAS, 457, 2377
\end{thebibliography}
\end{document}